\documentclass{article}

\usepackage{fullpage}
\usepackage{amsmath}
\usepackage{amssymb}
\usepackage{graphicx}
\usepackage{hyperref}
\usepackage{setspace}
\usepackage{authblk}

\begin{document}

\title{{\LARGE Some Relativistic and Gravitational Properties of the Wolfram Model}}
\author{{\Large Jonathan Gorard\footnote{\url{jg865@cam.ac.uk}}}}

\maketitle

\newtheorem{definition}{Definition}

\begin{abstract}
The Wolfram Model, which is a slight generalization of the model first introduced by Stephen Wolfram in \textit{A New Kind of Science} (NKS), is a discrete spacetime formalism in which space is represented by a hypergraph whose dynamics are determined by abstract replacement operations on set systems, and in which the conformal structure of spacetime is represented by a causal graph. The purpose of this article is to present rigorous mathematical derivations of many key properties of such models in the continuum limit, as first discussed in NKS, including the fact that large classes of them obey discrete forms of both special and general relativity. First, we prove that causal invariance (namely, the requirement that all causal graphs be isomorphic, irrespective of the choice of hypergraph updating order) is equivalent to a discrete version of general covariance, with changes to the updating order corresponding to discrete gauge transformations. This fact then allows one to deduce a discrete analog of Lorentz covariance, and the resultant physical consequences of discrete Lorentz transformations. We also introduce discrete notions of Riemann and Ricci curvature for hypergraphs, and prove that the correction factor for the volume of a discrete spacetime cone in a causal graph corresponding to curved spacetime of fixed dimensionality is proportional to a timelike projection of the discrete spacetime Ricci tensor, subsequently using this fact (along with the assumption that the updating rules preserve the dimensionality of the causal graph in limiting cases) to prove that the most general set of constraints on the discrete spacetime Ricci tensor corresponds to a discrete form of the Einstein field equations. Finally, we discuss a potential formalism for general relativity in hypergraphs of varying local dimensionality, and the implications that this formalism may have for inflationary cosmology and the value of the cosmological constant. Connections to many other fields of mathematics - including mathematical logic, abstract rewriting systems, ${\lambda}$-calculus, category theory, automated theorem-proving, measure theory, fluid mechanics, algebraic topology and geometric group theory - are also discussed.
\end{abstract}

\section{Introduction}
\label{sec:CurrentStatus0}

Perhaps the single most significant idea conveyed within Stephen Wolfram's \textit{A New Kind of Science}\cite{wolfram}, and the initial intellectual seedling from which the contents of the book subsequently grow, is the abstract empirical discovery that the ``computational universe'' - that is, the space of all possible programs - is far richer, more diverse and more vibrant than one might reasonably expect. The fact that such intricate and complex behavior can be exhibited by computational rules as apparently elementary as the Rule 30\cite{wolfram2}\cite{wolfram3} and Rule 110\cite{cook}\cite{cook2} cellular automata, which are so straightforward to represent that they can easily be discovered by systematic enumeration, is profoundly counterintuitive to many people.

However, once one has truly absorbed and internalized this realization, it leads to an exceedingly tantalizing possibility: that perhaps, lying somewhere out there in the computational universe, is the rule for our physical universe\cite{wolfram4}. If an entity as remarkable as Rule 30 could be found just by an exhaustive search, then perhaps so too can a theory of fundamental physics. The idea that there could exist some elementary computational rule that successfully reproduces the entirety of the physical universe at first seems somewhat absurd, although there does not appear to be any fundamental reason (neither in physics, nor mathematics, nor philosophy) to presume that such a rule could \textit{not} exist. Moreover, if there is even a remote possibility that such a rule \textit{could} exist, then it's slightly embarrassing for us not to be looking for it. The objective of the Wolfram Physics Project is to enact this search.

Conducting such a search is far from straightforward, since (more or less by definition) any reasonable candidate for our universe will inevitably be rife with computational irreducibility. Definite lower bounds on the degree of computational irreducibility in the laws of physics were first conjectured by Wolfram\cite{wolfram5}, and one such bound was proved recently by the author\cite{gorard}, with the corollary that there exists an infinite class of classical physical experiments with undecidable observables. Furthermore, explicit complexity-theoretic bounds have also been recently demonstrated by the author for the case of a discrete spacetime formalism, in collaboration with Shah\cite{shah}.

\begin{figure}[ht]
\centering
\includegraphics[width=0.345\textwidth]{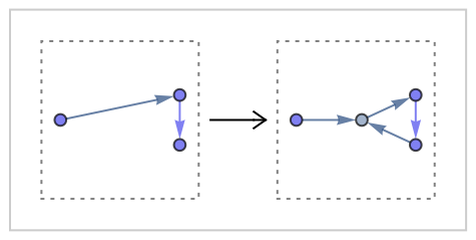}
\caption{An example of a possible replacement operation on a set system, visualized as a transformation rule between two hypergraphs (which, in this particular case, happen to be equivalent to ordinary graphs). Adapted from S. Wolfram, \textit{A Class of Models with Potential to Represent Fundamental Physics}\cite{wolfram_new}.}
\label{fig:CurrentStatus17}
\end{figure}

\begin{figure}[ht]
\centering
\includegraphics[width=0.445\textwidth]{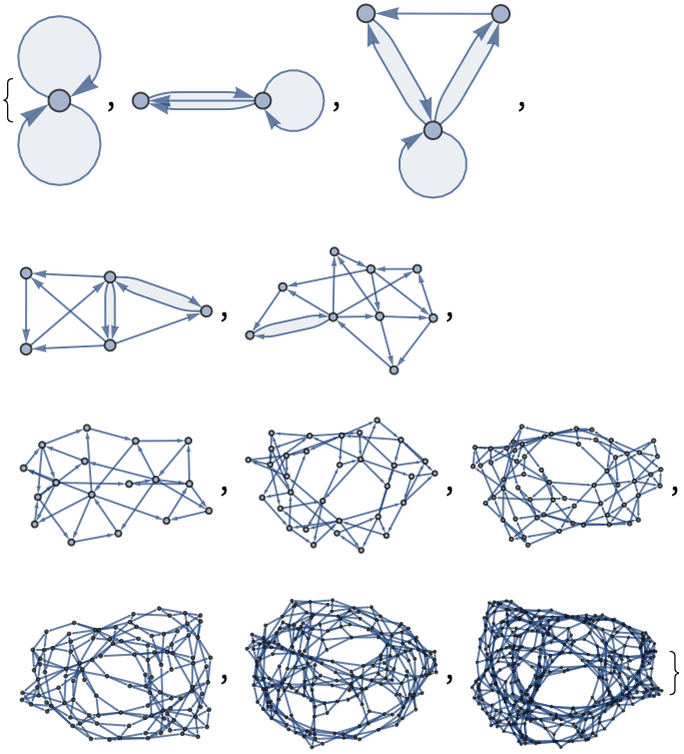}
\caption{An example evolution of the above transformation rule, starting from an initial (multi)hypergraph consisting of a single vertex with two self loops. Adapted from S. Wolfram, \textit{A Class of Models with Potential to Represent Fundamental Physics}.}
\label{fig:CurrentStatus18}
\end{figure}

\begin{figure}[ht]
\centering
\includegraphics[width=0.495\textwidth]{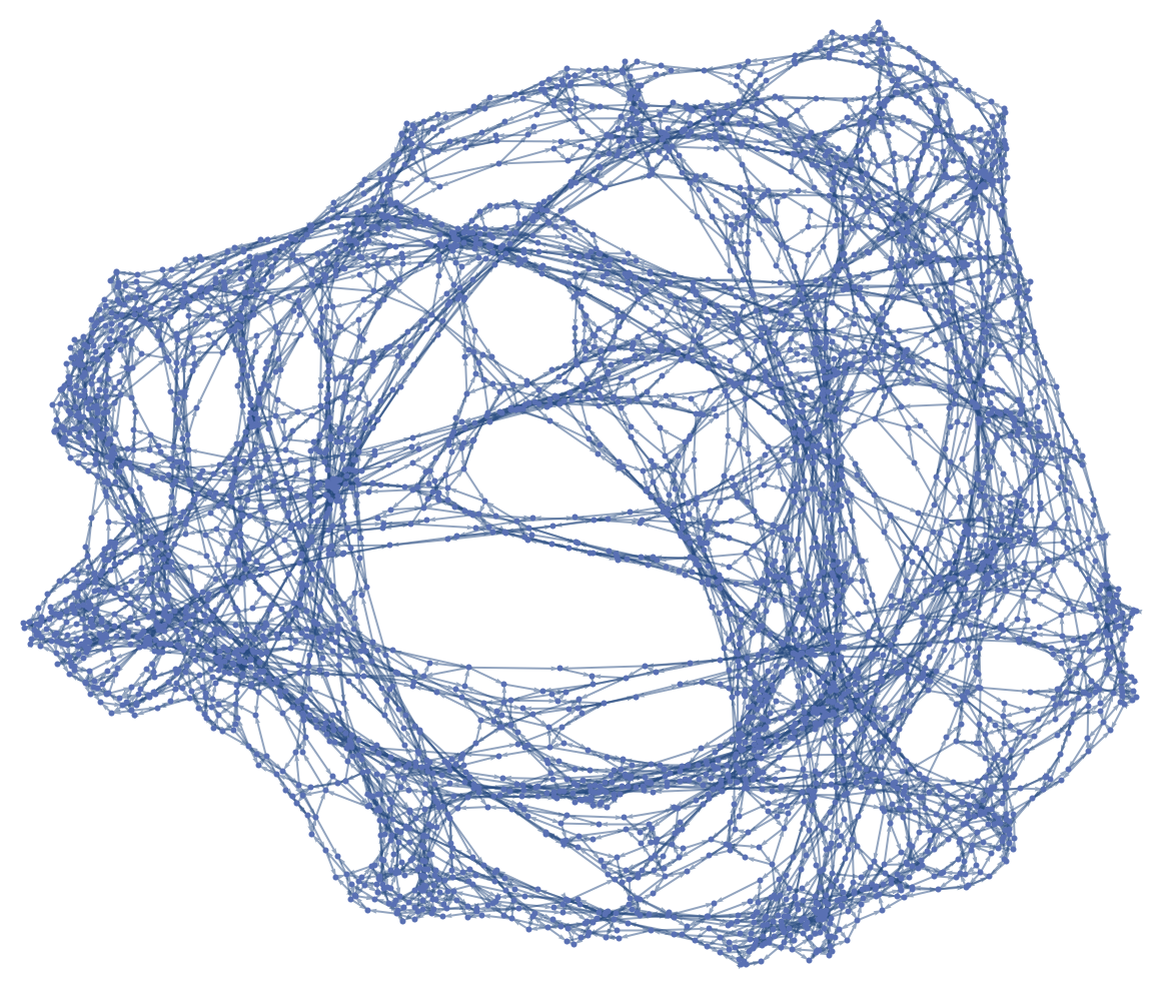}
\caption{The final state of the above Wolfram Model evolution. Adapted from S. Wolfram, \textit{A Class of Models with Potential to Represent Fundamental Physics}.}
\label{fig:CurrentStatus19}
\end{figure}

If one wants to construct a minimal model for the physical universe, then a cellular automaton is clearly an inappropriate formalism, as it already contains far too much inbuilt structure (such as a rigid array of cells, an explicit separation of the notion of space from the notion of states of cells, etc.). In NKS, Wolfram argued that a more suitable model for space would be a collection of discrete points, with connections defined between them in the form of a trivalent graph; laws of physics can then be represented by transformation rules defined on this graph (i.e. rules which replace subgraphs with other subgraphs, in such a way as to preserve the total number of outgoing edges). More recently, we have discovered that a slightly more general approach is to model space as a hypergraph, and hence to define transformation rules as replacement operations on set systems; an example of such a transformation rule is shown in Figure \ref{fig:CurrentStatus17}, and an example of its evolution is shown in Figures \ref{fig:CurrentStatus18} and \ref{fig:CurrentStatus19}.

The present article begins by outlining a basic mathematical definition of the Wolfram Model and its causal structure (in terms of discrete causal graphs) in Section \ref{sec:CurrentStatus1}, before proceeding to formalize these definitions using the theory of abstract rewriting systems from mathematical logic in Section \ref{sec:CurrentStatus2}, and discussing some connections to ${\lambda}$-calculus and category theory. Section \ref{sec:CurrentStatus3} builds upon these formalizations to develop a proof that causal invariance (namely, the requirement that all causal graphs be isomorphic, irrespective of the choice of hypergraph updating order) is equivalent to a discrete form of general covariance, with changes to the updating order corresponding to discrete gauge transformations. This fact allows one to deduce a discrete analog of Lorentz covariance, and the resultant physical consequences of discrete Lorentz transformations, as demonstrated in Section \ref{sec:CurrentStatus4}, using techniques from the theory of automated theorem-proving.

In Section \ref{sec:CurrentStatus5}, we introduce a discrete analog of the Ricci scalar curvature (namely the Ollivier-Ricci curvature for metric-measure spaces) for hypergraphs, and prove that the correction factor for the volume of a discrete ball in a spatial hypergraph corresponding to curved space of fixed dimensionality is trivially proportional to the Ollivier-Ricci curvature. We subsequently introduce discrete notions of parallel transport, holonomy, sectional curvature, the Riemann curvature tensor and the spacetime Ricci curvature tensor in Section \ref{sec:CurrentStatus6}, and use these notions to prove the corresponding result for volumes of discrete spacetime cones in causal graphs corresponding to curved spacetimes of fixed dimensionality, demonstrating that the correction factors are now proportional to timelike projections of the discrete spacetime Ricci tensor. In Section \ref{sec:CurrentStatus7}, we use this fact, along with the assumption that the updating rules preserve the dimensionality of the causal graph in limiting cases, to prove that the most general set of constraints on the discrete spacetime Ricci tensor correspond to a discrete form of the Einstein field equations, using an analogy to the Chapman-Enskog derivation of the continuum hydrodynamics equations from discrete molecular dynamics. Finally, in Section \ref{sec:CurrentStatus8}, we discuss a potential formalism for general relativity in hypergraphs of varying local dimensionality, using techniques from algebraic topology and geometric group theory, and the implications that this formalism may have for inflationary cosmology and the value of the cosmological constant.

\section{Space, Time and Causality in the Wolfram Model}

\subsection{Basic Formalism and Causal Graphs}
\label{sec:CurrentStatus1}

A much more complete description of this model is given by Wolfram in \cite{wolfram_new}. The essential idea here is to model space as a large collection of discrete points which, on a sufficiently large scale, resembles continuous space, in much the same way as a large collection of discrete molecules resembles a continuous fluid. A geometry can then be induced on this collection of points by introducing patterns of connections between them, as defined, for instance, by a hypergraph or set system. Here, a hypergraph denotes a direct generalization of an ordinary graph, in which (hyper)edges can join an arbitrary number of vertices\cite{berge}:

\begin{definition}
A ``spatial hypergraph'', denoted ${H = \left( V, E \right)}$, is a finite, undirected hypergraph, i.e:

\begin{equation}
E \subseteq \mathcal{P} \left( V \right) \setminus \lbrace \emptyset \rbrace,
\end{equation}
where ${\mathcal{P} \left( V \right)}$ denotes the power set of $V$.
\end{definition}
We assume henceforth that all hypergraphs are actually multihypergraphs, in the sense that $E$ is actually a multiset, thus allowing for hyperedges of arbitrary multiplicity.

Note that, for many practical purposes, it suffices to consider a special case of the more general hypergraph formulation, in which space is simply represented by a trivalent graph:

\begin{definition}
A ``trivalent spatial graph'', denoted ${G = \left( V, E \right)}$, is a finite, undirected, regular graph of degree 3.
\end{definition}
This was the approach adopted in NKS, and it suffices as a minimal representation for all finite, undirected graphs, because any vertex of degree greater than 3 could equivalently be replaced by a cycle of vertices of degree exactly 3, without changing any of the large-scale combinatorial properties of the graph.

The update rules, or ``laws of physics'', that effectively determine a particular candidate universe (up to initial conditions) are then defined to be abstract rewrite operations acting on these spatial hypergraphs, i.e. operations which take a subhypergraph with a particular canonical form, and replace it with a distinct subhypergraph with a different canonical form:

\begin{definition}
An ``update rule'', denoted $R$, for a spatial hypergraph ${H = \left( V, E \right)}$ is an abstract rewrite rule of the form ${H_1 \to H_2}$, in which a subhypergraph ${H_1}$ is replaced by a distinct subhypergraph ${H_2}$ with the same number of outgoing hyperedges as ${H_1}$, and in such a way as to preserve any symmetries of ${H_1}$.
\end{definition}

With the dynamics of the hypergraph thus defined, we are able to introduce something akin to the causal structure of a spacetime/Lorentzian manifold by constructing a directed graph in which every vertex corresponds to a spacetime ``event'' (i.e. an application of an update rule), and every edge specifies a causal relation between events.

\begin{definition}
A ``causal graph'', denoted ${G_{causal}}$, is a directed, acyclic graph in which every vertex corresponds to an application of an update rule (i.e. an ``event''), and in which the edge ${A \to B}$ exists if and only if the update rule designated by event $B$ was only applicable as a result of the outcome of the update rule designated by event $A$.
\end{definition}
Pragmatically speaking, this implies that the causal relation ${A \to B}$ exists if and only if the input for event $B$ has a non-trivial overlap with the output of event $A$. If the Wolfram Model is to be a plausible underlying formalism for fundamental physics, one must presumably assume that every edge of the causal graph corresponds to a spacetime interval on the order of (at most) the Planck scale, i.e. approximately ${10^{-35}}$ metres, or ${10^{-43}}$ seconds, although dimensional analysis of the model indicates a potentially much smaller spacetime scale\cite{wolfram_new}.

\begin{figure}[ht]
\centering
\includegraphics[width=0.495\textwidth]{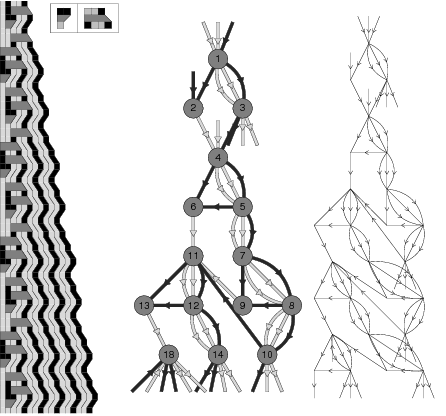}
\caption{A causal graph corresponding to the evolution of the string substitution system ${\lbrace BB \to A, AAB \to BAAB \rbrace}$, starting from the initial condition ${ABAAB}$. Adapted from S. Wolfram, \textit{A New Kind of Science}, page 498.}
\label{fig:CurrentStatus1}
\end{figure}

A causal graph corresponding to the evolution of an elementary string substitution system is shown in Figure \ref{fig:CurrentStatus1}.

In this respect, the Wolfram Model can be thought of as being an abstract generalization of the so-called ``Causal Dynamical Triangulation'' approach to quantum gravity developed by Loll, Ambj{\o}rn and Jurkiewicz\cite{loll}\cite{ambjorn}\cite{loll2}\cite{ambjorn2}, in which spacetime is topologically triangulated into a simplicial complex of 4-simplices (also known as pentachora), which then evolve according to some deterministic dynamical law.

\begin{figure}[ht]
\centering
\includegraphics[width=0.495\textwidth]{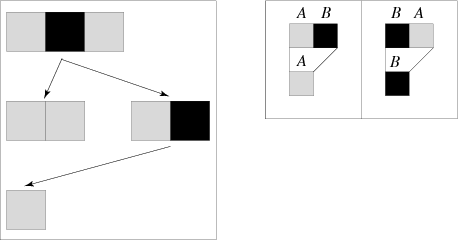}
\caption{A multiway system corresponding to the evolution of an elementary string substitution system ${\lbrace AB \to A, BA \to B \rbrace}$, starting from the initial condition ${ABA}$. Adapted from S. Wolfram, \textit{A New Kind of Science}, page 205.}
\label{fig:CurrentStatus16}
\end{figure}

With subhypergraph replacement rules of the general form ${H_1 \to H_2}$, it is evident that the evolution of a given spatial hypergraph will, in general, be nondeterministic, since there does not exist any canonical order in which the replacement rules should be applied, and distinct orderings will generally give rise to distinct spatial hypergraphs. In other words, the evolution history for an arbitrary candidate universe will, in most cases, correspond to a directed acyclic graph (known as a ``multiway system''), rather than a single path. A simple example of such a multiway system, corresponding to the evolution of an elementary string substitution system, is shown in Figure \ref{fig:CurrentStatus16}.

However, it turns out that there exist many situations in which one is able to mitigate this problem, and effectively obtain deterministic evolution of the hypergraph, by considering only replacement rules that exhibit a particular abstract property known as ``confluence'', as described below.

\subsection{Abstract Rewriting, the Church-Rosser Property and Causal Invariance}
\label{sec:CurrentStatus2}

Firstly, one is able to make the notions of update rules and transformations between hypergraphs more mathematically rigorous, by drawing upon the formalism of abstract rewriting systems in mathematical logic\cite{baader}\cite{bezem}.

\begin{definition}
An ``abstract rewriting system'' is a set, denoted $A$ (where each element of $A$ is known as an ``object''), equipped with some binary relation, denoted ${\to}$, known as the ``rewrite relation''.
\end{definition}
Concretely, ${a \to b}$ designates a replacement rule, indicating that object $a$ can be replaced with (or rewritten as) object $b$.

More generally, there may exist situations in which two objects, $a$ and $b$, are connected not by a single rewrite operation, but by some finite sequence of rewrite operations, i.e:

\begin{equation}
a \to a^{\prime} \to a^{\prime\prime} \to \dots \to b^{\prime} \to b,
\end{equation}
in  which case we can use the notation ${a \to^{*} b}$ to indicate the existence of such a rewrite sequence. More formally:

\begin{definition}
${\to^{*}}$ is the reflexive transitive closure of ${\to}$, i.e. it is the transitive closure of ${\to \cup =}$, where $=$ denotes the identity relation.
\end{definition}
In other words, ${\to^{*}}$ is the smallest preorder that contains ${\to}$, or the smallest binary relation that contains ${\to}$ and also satisfies both reflexivity and transitivity:

\begin{equation}
a \to^{*} a, \qquad \text{ and } \qquad a \to^{*} b, b \to^{*} c \implies a \to^{*} c.
\end{equation}

The concept of ``confluence'' allows us to formalize the idea that some objects in an abstract rewriting system may be rewritten in multiple ways, so as to yield the same eventual result:

\begin{definition}
An object ${a \in A}$ is ``confluent'', if:

\begin{equation}
\forall b, c \in A, \text{ such that } a \to^{*} b \text{ and } a \to^{*} c, \qquad \exists d \in A \text{ such that } b \to^{*} d \text{ and } c \to^{*} d.
\end{equation}
\end{definition}

\begin{figure}[ht]
\centering
\includegraphics[width=0.495\textwidth]{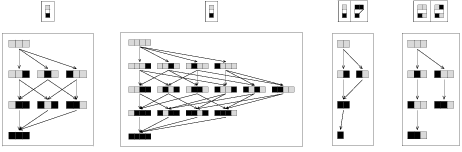}
\caption{Four evolution histories for (globally) confluent elementary string substitution systems (the first two corresponding to ${A \to B}$, and the last two to ${\lbrace A \to B, BB \to B \rbrace}$ and ${\lbrace AA \to BA, AB \to BA \rbrace}$, respectively) demonstrating that, irrespective of the chosen rewriting order, the same eventual result is always obtained. Adapted from S. Wolfram, \textit{A New Kind of Science}, pages 507 and 1037.}
\label{fig:CurrentStatus15}
\end{figure}

\begin{definition}
An abstract rewriting system $A$ is (globally) ``confluent'' (or exhibits the ``Church-Rosser property'') if every object ${a \in A}$ is confluent.
\end{definition}
Thus, within a (globally) confluent rewriting system, every time there exists an ambiguity in the rewriting order, such that distinct objects $b$ and $c$ can be obtained by different rewrite sequences from some common object $a$, those objects can always be made to reconverge on some common object $d$ after a finite number of rewriting operations. (Global) confluence is demonstrated explicitly for four elementary string substitution systems in Figure \ref{fig:CurrentStatus15}.

\begin{figure}[ht]
\centering
\includegraphics[width=0.495\textwidth]{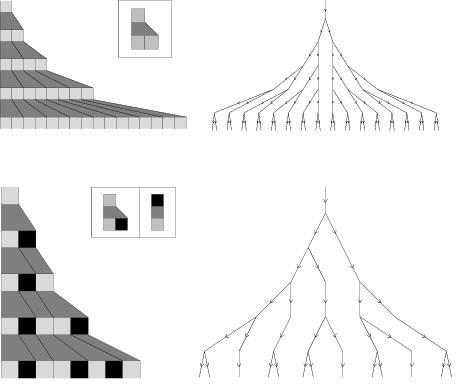}
\caption{A pair of nontrivial string substitution systems exhibiting causal invariance (${A \to AA}$ and ${\lbrace A \to AB, B \to A \rbrace}$, respectively), since in both cases the combinatorial structure of the causal graph is independent of the order in which the string substitutions get applied. Adapted from S. Wolfram, \textit{A New Kind of Science}, page 500.}
\label{fig:CurrentStatus2}
\end{figure}

The notion of confluence hence allows us to assuage the reader's potential fears regarding the potentially ambiguous ordering of update operations on spatial hypergraphs. If we consider only confluent hypergraph replacement rules, then, for any apparent divergence in the spatial hypergraphs that one obtains by following different paths of the multiway system (i.e. any divergence resulting from ambiguity in the ordering of the update operations), there will always exist a new path in the multiway system (i.e. a new sequence of update operations) which causes that ambiguity to disappear. We can formalize this notion by stating that confluence is a necessary condition for such rules to exhibit an asymptotic property known as ``causal invariance'':

\begin{definition}
A multiway system is ``causal invariant'' if the causal graphs that it generates (i.e. the causal graphs generated by following every possible path in the multiway graph, corresponding to every possible updating order) are all, eventually, isomorphic as directed, acyclic graphs.
\end{definition}
A pair of simple but nontrivial string substitution systems exhibiting causal invariance is shown in Figure \ref{fig:CurrentStatus2}.

There are some subtleties here, since the definition of (global) confluence only guarantees that the rewrite sequences ${b \to^{*} d}$ and ${c \to^{*} d}$ must exist for a given ${b, c}$, but it does not guarantee anything about their potential length. As such, the paths that one must follow in order to obtain convergence may be arbitrarily long, so although causal invariance necessitates that the causal graphs generated by following every path through the multiway system must \textit{eventually} become isomorphic, those causal graphs are not guaranteed to be isomorphic after any finite number of update steps. As such, causal invariance is best interpreted as a limiting statement about the global structure of the multiway system.

Outside of the theory of abstract rewriting systems, there are many equivalent ways to formalize these notions of update rules, confluence, causal invariance, etc. For instance, one can formulate a category-theoretic version of the same ideas, by first considering the rewrite relation ${\to}$ of the abstract rewriting system ${\left( A, \to \right)}$ to be an indexed union of subrelations:

\begin{equation}
\to_{1} \cup \to_{2} = \to.
\end{equation}
This system is mathematically equivalent to a labeled state transition system, ${\left( A, \Lambda, \to \right)}$, with ${\Lambda}$ being the set of indices (labels), and this system is itself simply a bijective function from $A$ to the powerset of $A$ indexed by ${\Lambda}$, i.e. ${\mathcal{P} \left( \Lambda \times A \right)}$, given by:

\begin{equation}
p \mapsto \lbrace \left( \alpha, q \right) \in \Lambda \times A : p \to^{\alpha} q \rbrace.
\end{equation}
Now, using the fact that the power set construction on the category of sets is a covariant endofunctor, the state transition system is an F-coalgebra for the functor ${\mathcal{P} \left( \Lambda \times \left( - \right) \right)}$. Then, the more general case in which ${\to}$ is not an indexed union of subrelations, which corresponds to an unlabeled state transition system, is simply the case in which ${\Lambda}$ is a singleton set. Therefore, in general, if ${\mathcal{P}}$ is considered to be an endofunctor on the category ${Set}$:

\begin{equation}
\mathcal{P} : Set \mapsto Set,
\end{equation}
then the system ${\left( A, \to \right)}$ is the object $A$ equipped with a morphism of ${Set}$, denoted ${\to}$:

\begin{equation}
\to : A \mapsto \mathcal{P} A.
\end{equation}
Statements about confluence and convergences between states in the multiway system thus translate into purely category-theoretic statements regarding limits of functors and the relationship between cones and co-cones (which can be thought of as being, respectively, the category-theoretic analogs of the notions of critical pair divergence and convergence in abstract rewriting theory)\cite{stokkermans}.

There exist certain weakenings of the property of (global) confluence, including ``local confluence'', in which only objects which diverge after a single rewrite operation are required to reconverge:

\begin{equation}
\forall b, c \in A, \text{ such that } a \to b \text{ and } a \to c, \qquad \exists d \in A \text{ such that } b \to^{*} d \text{ and } c \to^{*} d,
\end{equation}
and ``semi-confluence'', whereby one object is obtained by a single rewrite operation and the other is obtained by an arbitrary rewrite sequence:

\begin{equation}
\forall b, c \in A, \text{ such that } a \to b \text{ and } a \to^{*} c, \qquad \exists d \in A \text{ such that } b \to^{*} d \text{ and } c \to^{*} d.
\end{equation}
There are also various strengthenings, such as the strong ``diamond property'', in which objects that diverge after a single rewrite operation are also required to reconverge after a single rewrite operation:

\begin{equation}
\forall b, c \in A, \text{ such that } a \to b \text{ and } a \to c, \qquad \exists d \in A \text{ such that } b \to d \text{ and } c \to d,
\end{equation}
and ``strong confluence'', whereby, for two objects which diverged after a single rewrite operation, one is required to converge with a single rewrite operation (or no rewrite operations at all), whilst the other can converge via an arbitrary rewrite sequence:

\begin{equation}
\forall b, c \in A, \text{ such that } a \to b \text{ and } a \to c, \qquad \exists d \in A \text{ such that } b \to^{*} d \text{ and } \left( c \to d \text{ or } c = d \right).
\end{equation}
These properties, and their relation to the critical pair lemma in mathematical logic, are believed to be of foundational relevance to the derivation of quantum mechanics and quantum field theory within the Wolfram Model, as outlined in the accompanying publication\cite{gorard_new}.

The property of global confluence is often referred to as the Church-Rosser property for largely historical reasons; Alonzo Church and J. Barkley Rosser proved in 1936\cite{church} that ${\beta}$-reduction of ${\lambda}$ terms in the ${\lambda}$-calculus is globally confluent. In other words, if one considers the operation of replacing bound variables in the body of an abstraction as being a rewrite operation of the form:

\begin{equation}
\left( \left( \lambda x . M \right) E \right) \to \left( M \left[ x := E \right] \right).
\end{equation}
then two ${\lambda}$ terms are equivalent if and only if they are joinable. Here, ``equivalence'' of terms $x$ and $y$ refers to the binary operation ${x \leftrightarrow^{*} y}$, which is the reflexive transitive symmetric closure of ${\to}$, or the smallest equivalence relation containing ${\to}$ (otherwise described as the transitive closure of ${\leftrightarrow \cup =}$, where ${\leftrightarrow}$ denotes the symmetric closure of ${\to}$, i.e. ${\to \cup \to^{-1}}$). Moreover, ``joinability'' refers to the binary operation ${x \downarrow y}$, which designates the existence of a common term $z$ to which $x$ and $y$ are both reducible:

\begin{equation}
\forall x, y, x \downarrow y \iff \exists z \text{ such that } x \to^{*} z \leftarrow^{*} y.
\end{equation}

Having shown that (global) confluence is a necessary condition for causal invariance, we now proceed to prove that causal invariance implies an appropriately discretized form of general covariance, from which one is able to deduce both special and general relativity (in the form of Lorentz covariance and local Lorentz covariance, respectively).

\subsection{Causal Graph Hypersurfaces and Discrete Lorentz Covariance}
\label{sec:CurrentStatus3}

The first essential step in the derivation of special relativity for causal-invariant Wolfram Model systems is to make precise the formal correspondence between directed edges connecting updating events in a discrete causal graph, and timelike-separation of events in a continuous Minkowski space (or, more generally, in a Lorentzian manifold). In doing this, it is helpful first to introduce a canonical method for laying out a causal graph in Euclidean space, which can be defined as an optimization problem of the following form\cite{battista}\cite{battista2}:

\begin{definition}
A ``layered graph embedding'' is an embedding of a directed, acyclic graph in the Euclidean plane, in which edges are represented as monotonic downwards curves, and in which crossings between edges are to be minimized.
\end{definition}
The ideal case of a layered graph embedding, which is not guaranteed to exist in general, would be a so-called ``downward planar embedding'':

\begin{definition}
A ``downward planar embedding'' is an embedding of a directed, acyclic graph in the Euclidean plane, in which edges are represented as monotonic downwards curves without crossings.
\end{definition}

In the standard mathematical formalism for special relativity\cite{naber}, one starts by considering the ${(n + 1)}$-dimensional Minkowski space ${\mathbb{R}^{1, n}}$, in which every point ${\mathbf{p}}$ is a spacetime event (specified by one time coordinate and $n$ spatial coordinates). If one now considers performing a layered graph embedding of a causal graph into the discrete ``Minkowski lattice'' ${\mathbb{Z}^{1, n}}$, then one can label updating events by:

\begin{equation}
\mathbf{p} = \left( t, \mathbf{x} \right),
\end{equation}
where ${t \in \mathbb{Z}}$ is a discrete time coordinate, and:

\begin{equation}
\mathbf{x} = \left( x_1, \dots, x_n \right) \in \mathbb{Z}^n,
\end{equation}
are discrete spatial coordinates. One can then induce a geometry on ${\mathbb{Z}^{1, n}}$ by making an appropriate choice of norm:

\begin{definition}
The ``discrete Minkowski norm'' is given by:

\begin{equation}
\lVert \left( t, \mathbf{x} \right) \rVert = \lVert \mathbf{x} \rVert^2 - t^2,
\end{equation}
or, in more explicit form:

\begin{equation}
\lVert \left( t, \mathbf{x} \right) \rVert = \left( x_{1}^{2} + \dots + x_{n}^{2} \right) - t^2.
\end{equation}
\end{definition}

\begin{definition}
Updating events ${\mathbf{p} = \left( t, \mathbf{x} \right)}$ are classified as either ``timelike'', ``lightlike'' or ``spacelike'' based upon their discrete Minkowski norm:

\begin{align}
\mathbf{p} \sim \begin{cases}
\text{timelike}, \qquad &\text{ if } \lVert \left( t, \mathbf{x} \right) \rVert < 0,\\
\text{lightlike}, \qquad &\text{ if } \lVert \left( t, \mathbf{x} \right) \rVert = 0,\\
\text{spacelike}, \qquad &\text{ if } \lVert \left( t, \mathbf{x} \right) \rVert > 0.
\end{cases}
\end{align}
\end{definition}

\begin{definition}
Pairs of updating events ${\mathbf{p} = \left( t_1, \mathbf{x}_1 \right)}$, ${\mathbf{q} = \left( t_2, \mathbf{x}_2 \right)}$ can be classified as either ``timelike-separated'', ``lightlike-separated'' or ``spacelike-separated'', accordingly:

\begin{align}
\left( \mathbf{p}, \mathbf{q} \right) \sim \begin{cases}
\text{timelike-separated}, \qquad &\text{ if } \left( \left( t_1, \mathbf{x}_1 \right) - \left( t_2, \mathbf{x}_2 \right) \right) \sim \text{timelike},\\
\text{lightlike-separated}, \qquad &\text{ if } \left( \left( t_1, \mathbf{x}_1 \right) - \left( t_2, \mathbf{x}_2 \right) \right) \sim \text{lightlike},\\
\text{spacelike-separated}, \qquad &\text{ if } \left( \left( t_1, \mathbf{x}_1 \right) - \left( t_2, \mathbf{x}_2 \right) \right) \sim \text{spacelike}.
\end{cases}
\end{align}
\end{definition}

One of the foundational features of conventional special relativity is that two events are causally related if and only if they are timelike-separated. From our definition of the discrete Minkowski norm and the properties of layered graph embedding, we can see that a pair of updating events are causally related (i.e. connected by a directed edge in the causal graph) if and only if the corresponding vertices are timelike-separated in the embedding of the causal graph into the discrete Minkowski lattice ${\mathbb{Z}^{1, n}}$, as required.

\begin{figure}[ht]
\centering
\includegraphics[width=0.495\textwidth]{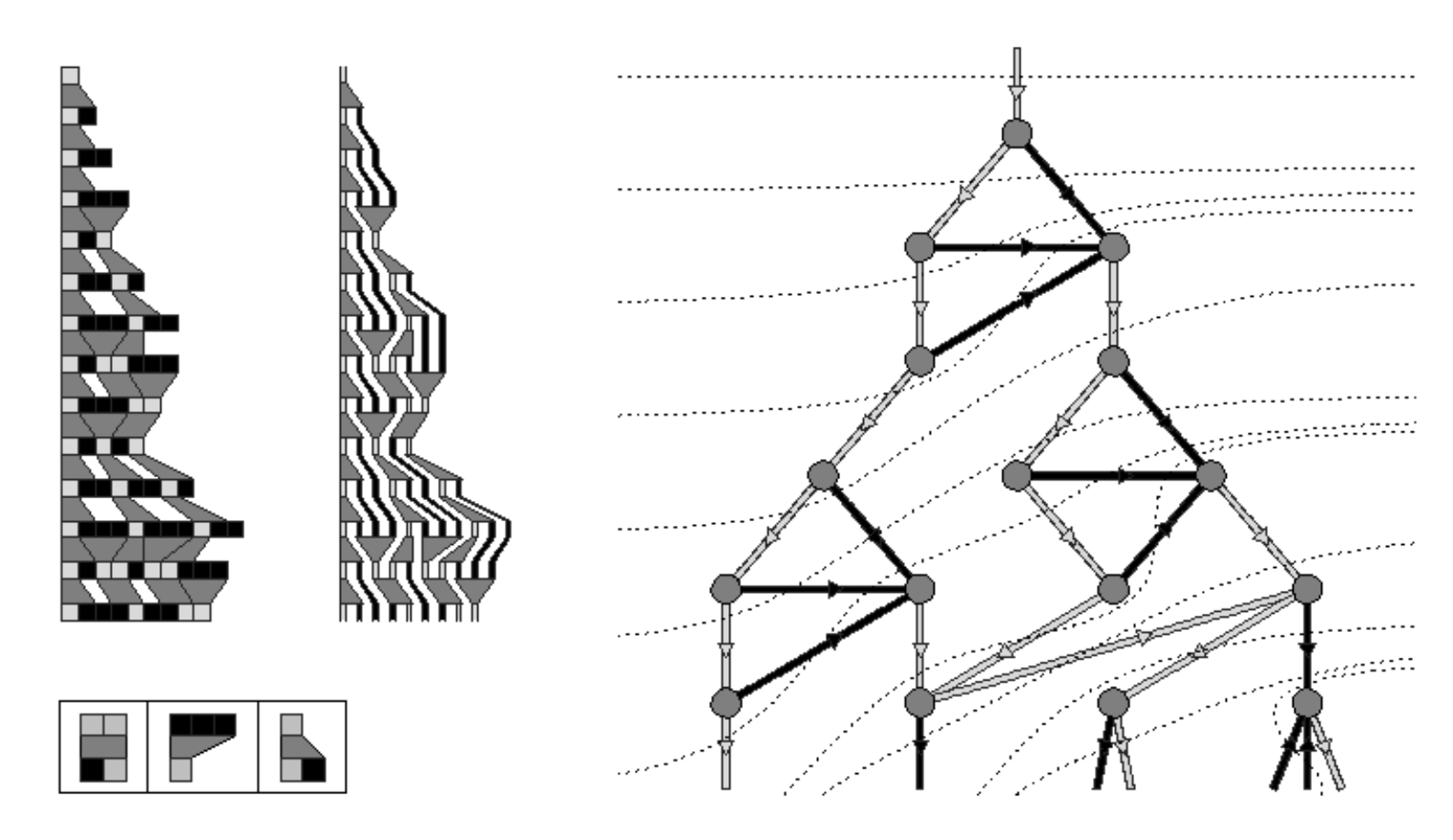}
\caption{A simple string substitution system (${AA \to BA}$, ${BBB \to A}$, ${A \to AB}$) that is not causal-invariant, since the causal graph is not unique. As a result, different possible discrete foliations of a given causal graph into discrete spacelike hypersurfaces (shown as dashed lines), corresponding to different possible choices of updating order for the substitution system, are found to produce distinct eventual outcomes for the system. Adapted from S. Wolfram, \textit{A New Kind of Science}, page 516.}
\label{fig:CurrentStatus3}
\end{figure}

Different possible layerings of the causal graph, or, equivalently, different possible ``slicings'' through the causal graph taken in the canonical layered graph embedding, will therefore correspond to different possible permutations in the ordering of the updating events. Every possible update scheme hence corresponds to a different possible discrete foliation of the causal graph into these ``slices'', where each ``slice'' effectively designates a possible spatial hypergraph that can be generated by some permutation of the updating events. Each such slice thus constitutes a discrete spacelike hypersurface, i.e. a discrete hypersurface embedded in ${\mathbb{Z}^{1, n}}$, in which every pair of updating events is spacelike-separated. A possible discrete hypersurface foliation (and hence, a possible updating order) of the causal graph for a non-causal-invariant string substitution system is shown in Figure \ref{fig:CurrentStatus3}.

Physically, each such slice through the causal graph is the discrete analog of a Cauchy surface in spacetime, which one can show by making an explicit correspondence with the so-called ``ADM decomposition'', developed by Arnowitt, Deser and Misner in 1959\cite{arnowitt}\cite{misner}, for Hamiltonian general relativity. In the ADM formalism, one decomposes a 4-dimensional spacetime (consisting of a manifold and a Lorentzian metric) by foliating it into a parameterized family of non-intersecting spacelike hypersurfaces; it is fairly straightforward to extend this notion to the case of discrete causal graphs:

\begin{definition}
A ``discrete spacetime'' is any pair ${\left( \mathcal{M}, g \right)}$, where ${\mathcal{M}}$ is a discrete metric space (taken to be the discrete analog of a pseudo-Riemannian manifold), and $g$ is a discrete Lorentzian metric of signature ${\left( - + + + \right)}$.
\end{definition}
(Note that the mathematical details of how such a discrete Lorentzian metric tensor can be explicitly obtained are outlined in the next section.)

\begin{definition}
A ``discrete Cauchy surface'' is a discrete spacelike hypersurface, i.e a set of updating events:

\begin{equation}
\Sigma \subset \mathcal{M},
\end{equation}
with the property that every timelike or lightlike (i.e. null) path through the causal graph, without endpoints (taken to be the discrete analog of a smooth timelike or lightlike curve), intersects an updating event in  ${\Sigma}$ exactly once .
\end{definition}

We proceed on the assumption that our spacetime is globally discretely hyperbolic, in the sense that there exists some universal ``time function'':

\begin{equation}
\exists t : \mathcal{M} \to \mathbb{Z},
\end{equation}
with non-zero gradient everywhere:

\begin{equation}
\mathbf{\Delta} t \neq 0 \text{ everywhere},
\end{equation}
such that our spacetime can be discretely foliated into non-intersecting level sets of this function, and that the collection of such level sets successfully covers the entire spacetime, i.e. the level surfaces ${t = const.}$ are discrete hypersurfaces, with the property:

\begin{equation}
\forall t_1, t_2 \in \mathbb{Z}, \Sigma_{t_1} = \lbrace p \in \mathcal{M} : t (p) = t_1 \rbrace, \text{ and } \Sigma_{t_1} \cap \Sigma_{t_2} = \emptyset \iff t_1 \neq t_2.
\end{equation}

Then, by correspondence with the standard ADM decomposition, the proper distances on each discrete hypersurface, denoted ${\Delta l}$, are determined by the induced 3-dimensional discrete spatial metric, denoted ${\gamma_{i j}}$:

\begin{equation}
\Delta l^2 = \gamma_{i j} \Delta x^i \Delta x^j,
\end{equation}
where the spatial coordinates ${x^i \left( t \right)}$ label the points on the discrete hypersurface, for a given coordinate time $t$. The normal direction to the discrete hypersurface is given locally by some vector, denoted ${\mathbf{n} \in \mathbb{Z}^{1, n}}$, representing the discrete relativistic 4-velocity of a normal observer; by travelling along the normal direction, the distance in proper time, denoted ${\Delta \tau}$, to the adjacent discrete hypersurface at time ${t + \Delta t}$ is given by:

\begin{equation}
\Delta \tau = \alpha \Delta t,
\end{equation}
where ${\alpha}$ is the ``lapse function'', i.e. a gauge variable that determines how the spacetime is discretely foliated in the timelike direction. Analogously, a set of three gauge variables, known collectively as the ``shift vector'', denoted ${\beta^i}$, determines how the spacetime is discretely foliated in the spacelike direction. The shift vectors effectively relabel the discrete spatial coordinates in accordance with the following scheme:

\begin{equation}
x^i \left( t + \Delta t \right) = x^i \left( t \right) - \beta^i \Delta t,
\end{equation}
such that the overall spacetime line element in our discrete analog of ADM becomes:

\begin{equation}
\Delta s^2 = \left( - \alpha^2 + \beta^i \beta_i \right) \Delta t^2 + 2 \beta_i \Delta x^i \Delta t + \gamma_{i j} \Delta x^i \Delta x^j.
\end{equation}

Thus, we can see that, under this formal analogy, the gauge freedom of the ADM formalism (i.e. our freedom to choose values of ${\alpha}$ and ${\beta^i}$ for each normal observer) corresponds directly to our freedom to choose an updating order, and hence a discrete foliation, for a given causal graph. In particular, ${\alpha}$ designates the effective number of updating events required to map between an input subhypergraph and the corresponding output subhypergraph in two neighboring hypersurfaces, and ${\beta^i}$ designates the effective graph distance between those corresponding subhypergraphs.

Note that, in the above, we have implicitly made a correspondence between the combinatorial structure of causal graphs, and the causal structure of Lorentzian manifolds\cite{penrose}\cite{levichev}\cite{malament}. In particular, we can exploit this correspondence to define notions of chronological and causal precedence for updating events, as follows:

\begin{definition}
An updating event $x$ ``chronologically precedes'' updating event $y$, denoted ${x \ll y}$, if there exists a future-directed (i.e. monotonic downwards) chronological (i.e. timelike) path through the causal graph connecting $x$ and $y$.
\end{definition}

\begin{definition}
An updating event $x$ ``strictly causally precedes'' updating event $y$, denoted ${x < y}$, if there exists a future-directed (i.e. monotonic downwards) causal (i.e. non-spacelike) path through the causal graph connecting $x$ and $y$.
\end{definition}

\begin{definition}
An updating event $x$ ``causally precedes'' updating event $y$, denoted ${x \prec y}$, if either $x$ strictly causally precedes $y$, or ${x = y}$.
\end{definition}
The standard algebraic properties of these relations, such as transitivity:

\begin{equation}
x \ll y, y \ll z \implies x \ll z, \qquad x \prec y, y \prec z \implies x \prec z,
\end{equation}
then follow by elementary combinatorics.

One can hence define the chronological and causal future and past for individual updating events:

\begin{definition}
The ``chronological future'' and ``chronological past'' of an updating event $x$, denoted ${I^{+} \left( x \right)}$ and ${I^{-} \left( x \right)}$, are defined as the sets of updating events which $x$ chronologically precedes, and which chronologically precede $x$, respectively:

\begin{equation}
I^{+} \left( x \right) = \lbrace y \in \mathcal{M} : x \ll y \rbrace, \qquad I^{-} \left( x \right) = \lbrace y \in \mathcal{M} : y \ll x \rbrace.
\end{equation}
\end{definition}

\begin{definition}
The ``causal future'' and ``causal past'' of an updating event $x$, denoted ${J^{+} \left( x \right)}$ and ${J^{-} \left( x \right)}$, are defined as the sets of updating events which $x$ causally precedes, and which casually precede $x$, respectively:

\begin{equation}
J^{+} \left( x \right) = \lbrace y \in \mathcal{M} : x \prec y \rbrace), \qquad J^{-} \left( x \right) = \lbrace y \in \mathcal{M} : y \prec x \rbrace.
\end{equation}
\end{definition}
For instance, in the discrete Minkowski space, ${\mathbb{Z}^{1, n}}$, considered above, ${I^{+} \left( x \right)}$ designates only the interior of the future light cone of $x$, whereas ${J^{+} \left( x \right)}$ designates the entire future light cone (i.e. including the cone itself). Chronological and causal future and past can accordingly be defined for sets of updating events, denoted ${S \subset \mathcal{M}}$, as follows:

\begin{equation}
I^{\pm} \left( S \right) = \bigcup_{x \in S} I^{\pm} \left( x \right), \qquad J^{\pm} \left( S \right) = \bigcup_{x \in S} J^{\pm} \left( x \right).
\end{equation}
Once again, standard algebraic properties of the chronological and causal future and past, such as the fact that the interiors of past and future light cones are always strict supersets of the light cones themselves:

\begin{equation}
I^{+} \left( S \right) = I^{+} \left( I^{+} \left( S \right) \right) \subset J^{+} \left( S \right) = J^{+} \left( J^{+} \left( S \right) \right),
\end{equation}
and:

\begin{equation}
I^{-} \left( S \right) = I^{-} \left( I^{-} \left( S \right) \right) \subset J^{-} \left( S \right) = I^{-} \left( I^{-} \left( S \right) \right),
\end{equation}
also hold combinatorially.

One rather welcome consequence of this new formalism is that it allows us to introduce a more rigorous definition of a discrete Cauchy surface within a causal graph (and hence, to provide a more complete mathematical description of a discrete causal graph foliation):

\begin{definition}
The ``discrete future Cauchy development'' of a set of updating events ${S \subset \mathcal{M}}$, denoted ${D^{+} \left( S \right)}$, is the set of all updating events $x$ for which every past-directed (i.e. monotonic upwards), inextendible, causal (i.e. non-spacelike) path in the causal graph through $x$ also intersects an updating event in $S$ at least once. Likewise for the ``discrete past Cauchy development'', denoted ${D^{-} \left( S \right)}$.
\end{definition}

\begin{definition}
The ``discrete Cauchy development'' of a set of updating events ${S \subset \mathcal{M}}$, denoted ${D \left( S \right)}$, is the union of future and past Cauchy developments:

\begin{equation}
D \left( S \right) = D^{+} \left( S \right) \cup D^{-} \left( S \right).
\end{equation}
\end{definition}

\begin{definition}
A set of updating events ${S \subset \mathcal{M}}$ is ``achronal'' if $S$ is disjoint from its own chronological future, i.e:

\begin{equation}
\nexists q, r \in S, \text{ such that } r \in I^{+} \left( q \right).
\end{equation}
\end{definition}

\begin{definition}
A ``discrete Cauchy surface'' in ${\mathcal{M}}$ is an achronal set of updating events whose discrete Cauchy development is ${\mathcal{M}}$.
\end{definition}
Another elegant byproduct of this formalism is that it makes manifest the connection between causal graphs and conformal transformations; specifically, the causal graph represents the conformally-invariant structure of a discrete Lorentzian manifold. Since the combinatorial structure of a causal graph is unchanged by its embedding, one can conclude that the causal structure of a discrete Lorentzian manifold is invariant under the conformal transformation:

\begin{equation}
\hat{g} = \Omega^2 g,
\end{equation}
for conformal factor ${\Omega}$, since the timelike, null and spacelike qualities of tangent vectors, denoted ${X}$, remain invariant under this map, e.g:

\begin{equation}
g \left( X, X \right) < 0, \qquad \implies \hat{g} \left( X, X \right) = \Omega^2 g \left( X, X \right) < 0.
\end{equation}

Therefore, the requirement of causal invariance for a Wolfram Model corresponds precisely to the claim that the ordering of timelike-separated updating events is agreed upon by all observers, irrespective of their particular choice of discrete spacelike hypersurface foliation (i.e. it is invariant under changes in the updating order of the system), even though the ordering of spacelike-separated events is not in general. In the most generic case, this can be interpreted as the claim that the eventual outcomes of updating events are independent of the reference frame (i.e. discrete hypersurface) in which the observer is embedded, which is a discretized version of the principle of general covariance, or diffeomorphism invariance, of the laws of physics.

In the particular case in which the discrete hypersurfaces contained within the foliation are required to be ``flat'' (a rigorous definition of curvature for a discrete hypersurface will be presented within the next section), those hypersurfaces will thus correspond to inertial reference frames, and the statement of causal invariance reduces to a statement of discrete Lorentz covariance. Since the other fundamental postulate of special relativity - namely constancy of the speed of light - is enforced axiomatically by our definition of the edge lengths in causal graphs, this completes the proof. More explicit details of how the physical consequences of discrete Lorentz transformations may be derived are outlined below.

It is worth noting that the true distinction between inertial and non-inertial frames in the Wolfram Model is ultimately a computability-theoretic one; one may think of an observer in relativity as being an entity which, upon observing the evolution of a collection of clocks, attempts to construct a hypothetical gravitational field configuration which is consistent with that evolution (i.e. to ``synchronize'' these clocks). Therefore, the relative computational power of the observer and the clocks places constraints upon the types of field configurations (i.e. reference frames) that the observer is able to construct within their internal model of the world. For instance, an observer who is unable to decide membership of any set of integers defined by a ${\Pi_1}$ or ${\Sigma_1}$ sentence in the arithmetical hierarchy must be unable to construct a ``Malament-Hogarth'' spacetime, ${\left( \mathcal{M}, g \right)}$, of the form:

\begin{equation}
\exists \gamma_1 \subset \mathcal{M} \text{ and } p \in \mathcal{M}, \qquad \text{ such that } \int_{\gamma_1} d \tau = \infty \text{ and } \gamma_1 \subset I^{-} \left( p \right),
\end{equation}
since such spacetimes are known to permit the construction of hypercomputers, i.e. computers that are able to solve recursively undecidable problems in finite time\cite{welch}\cite{hogarth}\cite{hogarth2}\cite{hogarth3}, thus contradicting the stated computational power of the observer. Therefore, an inertial reference frame corresponds to the limiting case of an arbitrarily computationally-bounded observer, and other constraints on the geometry of spacetime (and the resultant energy conditions regarding the matter content of that spacetime) follow from the observer's own position in the arithmetical hierarchy. A formal statement of this argument, and a derivation of its computability-theoretic and relativistic consequences, will be outlined in a subsequent publication.

Note also that causal disconnections in spacetime, such as event horizons and apparent horizons, are now represented as literal disconnections in the combinatorial structure of the causal graph (for instance, the problem of determining whether null rays can escape to future null infinity from a region of spacetime can now be represented as a concrete reachability problem for vertices of the causal graph). This observation is relevant both to our exploration of quantum mechanics in \cite{gorard_new}, and to the implications of the Wolfram Model for such problems as the black hole information paradox and the weak cosmic censorship hypothesis, which shall be explored in future work.

\subsection{Discrete Lorentz Transformations}
\label{sec:CurrentStatus4}

Consider now how an observer embedded within one inertial reference frame (i.e. flat discrete hypersurface), denoted $F$, measures discrete coordinates relative an observer embedded within a different inertial reference frame, denoted ${F^{\prime}}$, which we take to be moving at some constant velocity, ${\mathbf{v} \in \mathbb{Z}^n}$, relative to $F$. Let:

\begin{equation}
\mathbf{v} = \left( \tanh \left( \rho \right) \right) \mathbf{u},
\end{equation}
where ${\mathbf{u} \in \mathbb{Z}^n}$ is a vector representing the direction of motion, and:

\begin{equation}
\tanh \left( \rho \right) = \frac{\lVert \mathbf{v} \rVert}{\lVert \mathbf{u} \rVert} < 1,
\end{equation}
represents the magnitude, appropriately normalized. By convention, we shall denote the coordinates in $F$ by ${\left( t, \mathbf{x} \right)}$, and the coordinates in ${F^{\prime}}$ by ${\left( t^{\prime}, \mathbf{x}^{\prime} \right)}$, with the two coordinate systems synchronized such that:

\begin{equation}
t = t^{\prime} = 0,
\end{equation}
when the inertial reference frames initially coincide.

\begin{definition}
The ``discrete Lorentz transformation'' expresses the ${F^{\prime}}$ coordinates in terms of the $F$ coordinates as:

\begin{equation}
t^{\prime} = \left( \cosh \left( \rho \right) \right) t - \left( \sinh \left( \rho \right) \right) \mathbf{x} \cdot \mathbf{u},
\end{equation}
where ${\cdot}$ denotes the standard (``dot'') inner product of vectors in ${\mathbb{Z}^n}$.
\end{definition}

\begin{figure}[ht]
\centering
\includegraphics[width=0.495\textwidth]{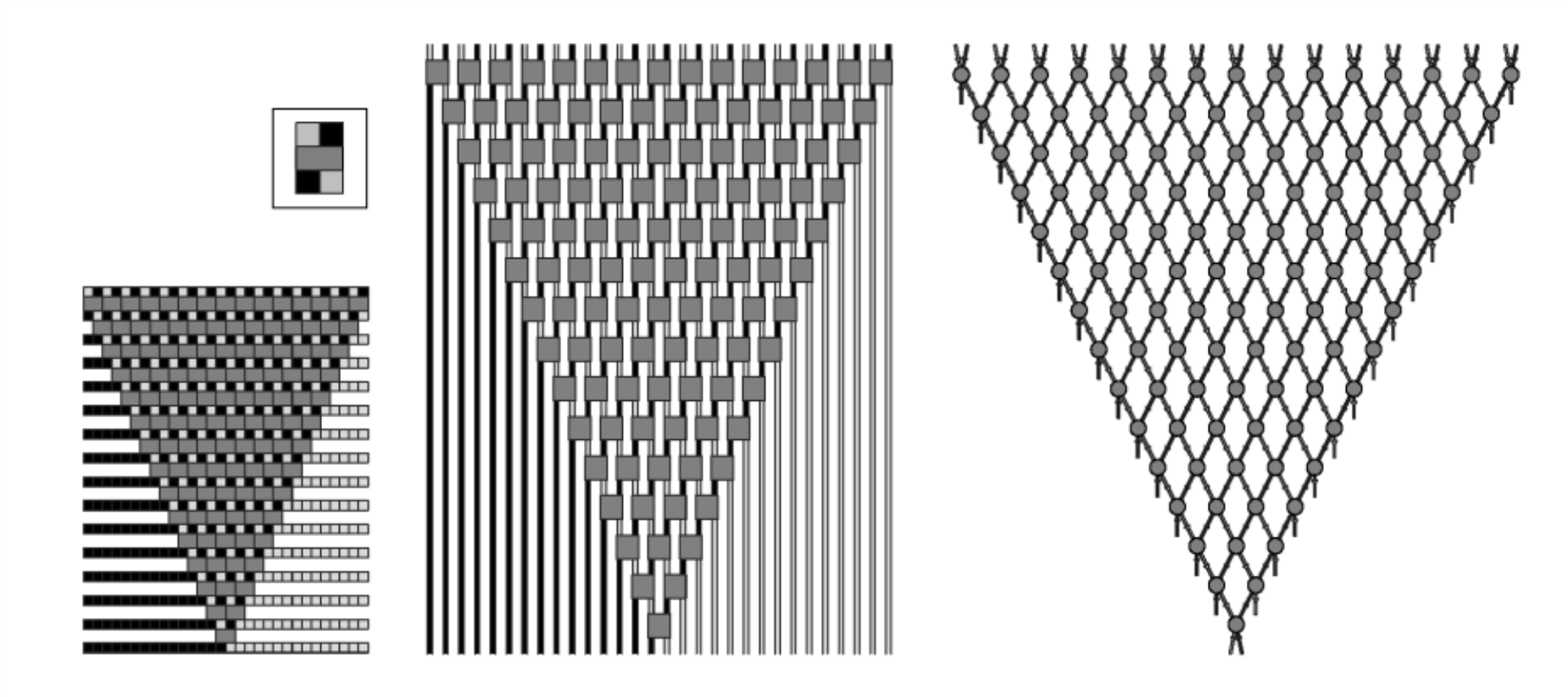}
\caption{An elementary and trivially causal-invariant string substitution system, ${AB \to BA}$, applied to the periodic initial condition ${ABABABAB \dots}$. Adapted from S. Wolfram, \textit{A New Kind of Science}, page 518.}
\label{fig:CurrentStatus4}
\end{figure}

To more clearly illustrate the connection between discrete Lorentz transformations and different choices of updating orders in a causal graph, consider the causal graph for the very simple (and trivially causal-invariant) string substitution system shown in Figure \ref{fig:CurrentStatus4}.

\begin{figure}[ht]
\centering
\includegraphics[width=0.495\textwidth]{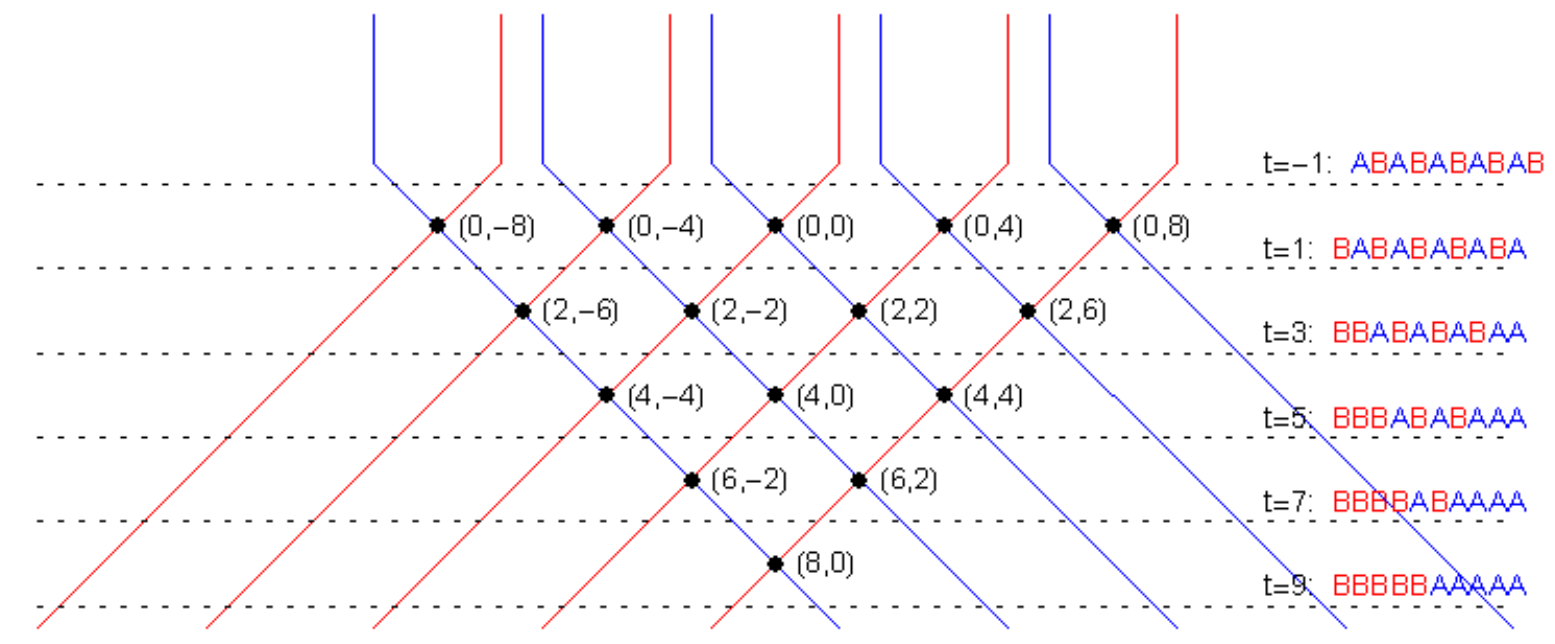}
\includegraphics[width=0.245\textwidth]{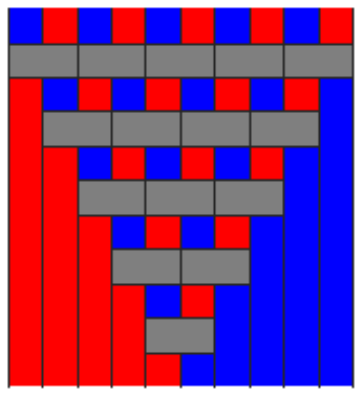}
\caption{On the left is the causal graph, with the coordinates for its standard layered graph embedding, corresponding to the default choice of updating order/discrete spacelike hypersurface foliation. On the right is the evolution history of the string substitution system for this default updating order (i.e. the updating order as seen by an observer in the rest frame $F$). Adapted from {\O}. Tafjord, \textit{NKS and the Nature of Space and Time}, slide 11.}
\label{fig:CurrentStatus5}
\end{figure}

The causal graph, with the coordinates for its default foliation into discrete spacelike hypersurfaces (as determined by the standard layered graph embedding), along with the corresponding updating order of the substitution system (which one can think of as being the ordering of updating events as seen by an observer in the rest frame $F$), are shown in Figure \ref{fig:CurrentStatus5}.

\begin{figure}[ht]
\centering
\includegraphics[width=0.495\textwidth]{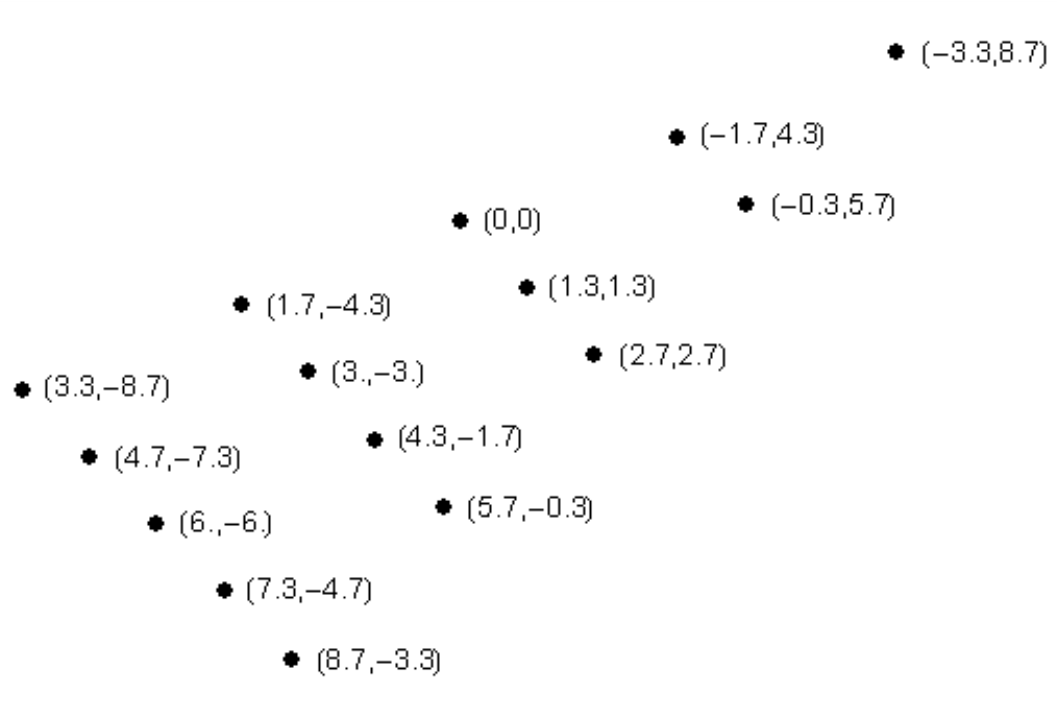}
\caption{The new, discretely Lorentz-transformed coordinates for the causal graph vertices (updating events) in the inertial frame ${F^{\prime}}$, moving with normalized velocity ${v = \frac{5}{13}}$ relative to the rest frame $F$. Adapted from {\O}. Tafjord, \textit{NKS and the Nature of Space and Time}, slide 12.}
\label{fig:CurrentStatus6}
\end{figure}

Now, applying a discrete Lorentz transformation allows us to consider a new inertial reference frame, ${F^{\prime}}$, presumed to be moving with a normalized velocity ${v = \frac{5}{13}}$ relative to the rest frame $F$. The vertex coordinates (i.e. the discrete spacetime coordinates of the updating events) corresponding to this new choice of causal graph embedding are shown in Figure \ref{fig:CurrentStatus6}.

\begin{figure}[ht]
\centering
\includegraphics[width=0.495\textwidth]{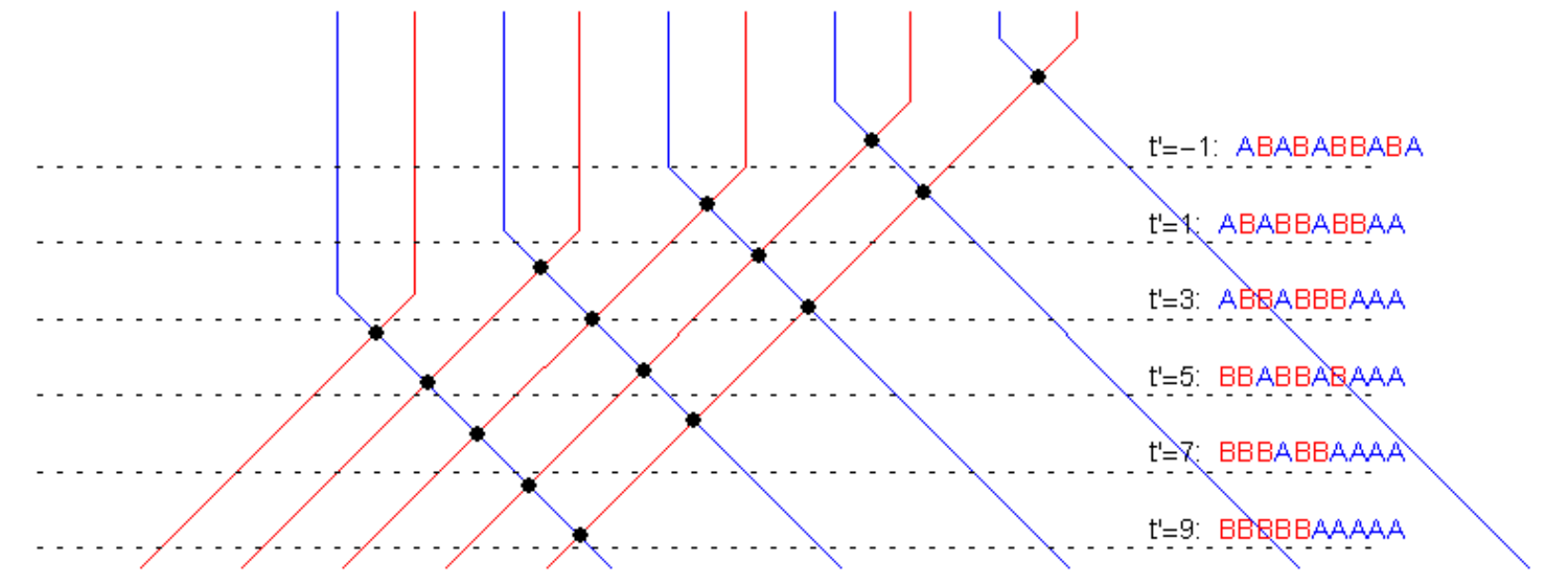}
\includegraphics[width=0.075\textwidth]{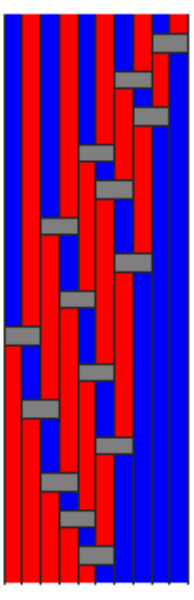}
\caption{On the left is the causal graph, with the new, discretely Lorentz-transformed updating order/hypersurface foliation shown. On the right is the evolution history of the string substitution system for this new choice of updating order (i.e. the updating order as seen by an observer in the inertial frame ${F^{\prime}}$). Adapted from {\O}. Tafjord, \textit{NKS and the Nature of Space and Time}, slide 13.}
\label{fig:CurrentStatus7}
\end{figure}

This discrete Lorentz transformation, leading to a new possible embedding of the causal graph, itself defines a new possible foliation of the causal graph into discrete spacelike hypersurfaces, and hence a new possible updating order for the substitution system. The new causal graph foliation and updating order are shown in Figure \ref{fig:CurrentStatus7}.

\begin{figure}[ht]
\centering
\includegraphics[width=0.495\textwidth]{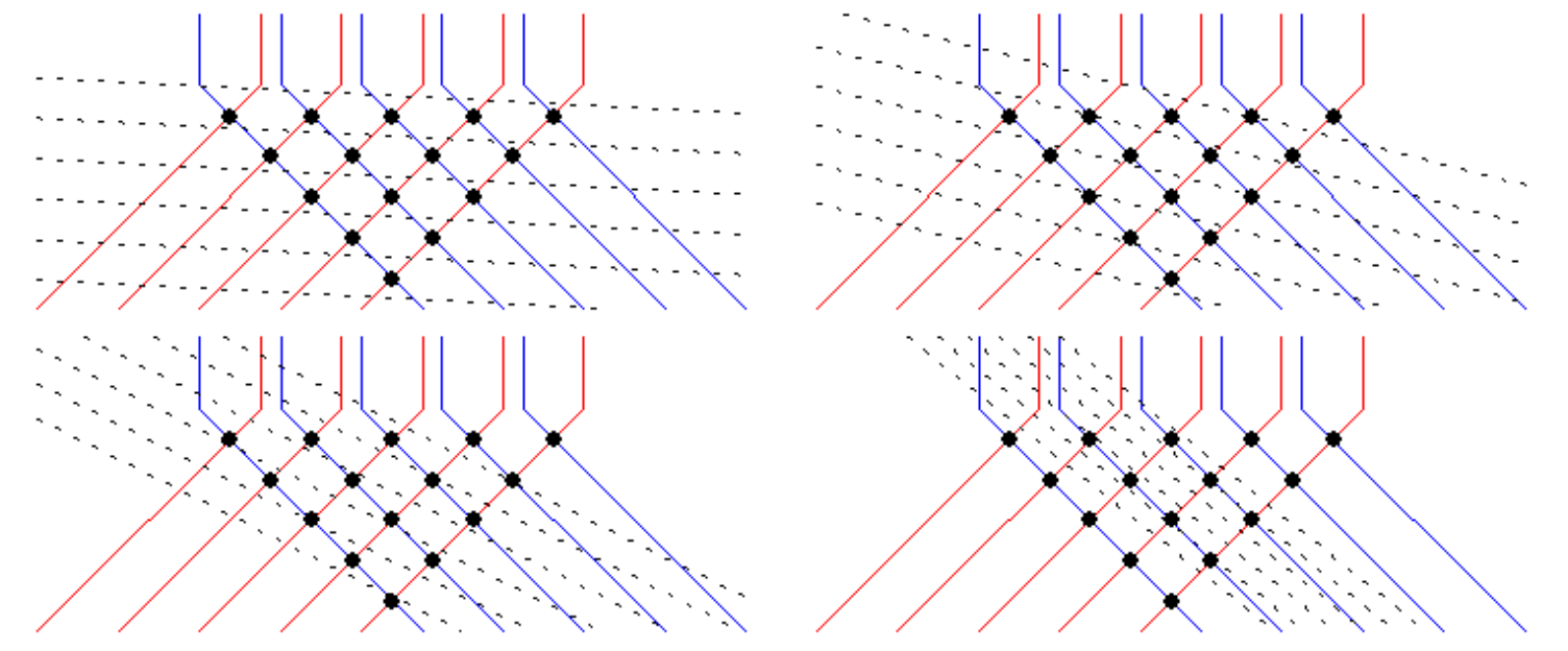}
\includegraphics[width=0.245\textwidth]{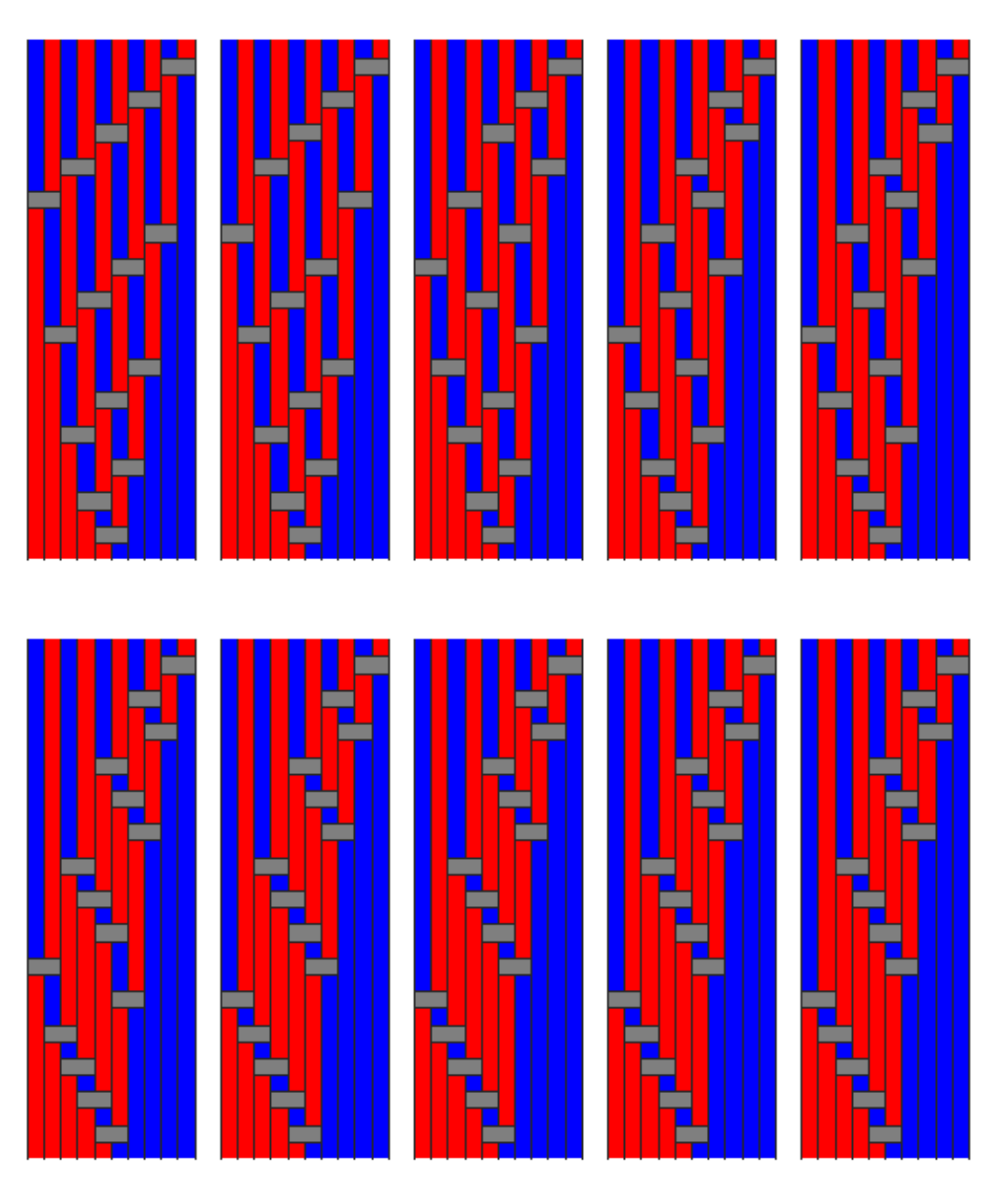}
\caption{On the left are the causal graphs, with discretely Lorentz-transformed hypersurface foliations corresponding to normalized velocities of ${v = 0.05}$, ${v = 0.25}$, ${v = 0.5}$ and ${v = 0.8}$, respectively. On the right are the evolution histories of the string substitution system, as seen by observers moving at normalized velocities of ${v = 0.05}$, ${v = 0.15}$, ${v = 0.25}$, ${v = 0.35}$, ${v = 0.45}$, ${v = 0.55}$, ${v = 0.65}$, ${v = 0.75}$, ${v = 0.85}$ and ${v = 0.95}$, respectively. Adapted from {\O}. Tafjord, \textit{NKS and the Nature of Space and Time}, slide 15.}
\label{fig:CurrentStatus8}
\end{figure}

As expected, we see that, following discrete Lorentz transformations corresponding to a variety of different (effectively subliminal) velocities, the ordering of spacelike-separated events may indeed vary, but the inherent causal invariance of the substitution rule guarantees that the ordering of timelike-separated events always remains invariant, as shown in Figure \ref{fig:CurrentStatus8}.

To understand the precise details of how concepts such as relativistic mass emerge within this new formalism, we must first introduce a notion of (elementary) particles in spatial hypergraphs. For the sake of simplicity, let us consider the particular case of ordinary graphs (i.e. hypergraphs in which each edge connects exactly two vertices). We now exploit a fundamental result in graph theory, known as ``Kuratowksi's theorem'', which states that a graph is planar (i.e. can be embedded in the Euclidean plane without any crossings of edges) if and only if it does not contain a subgraph that is a subdivision of either ${K_5}$ (the complete graph on 5 vertices), or ${K_{3, 3}}$ (the ``utility graph'', or bipartite complete graph on ${3 + 3}$ vertices)\cite{kuratowski}\cite{tutte}:

\begin{definition}
A ``subdivision'' of an undirected graph ${G = \left( V, E \right)}$ is a new, undirected graph ${H = \left( W, F \right)}$ which results from the subdivision of edges in $G$.
\end{definition}

\begin{figure}[ht]
\centering
\includegraphics[width=0.245\textwidth]{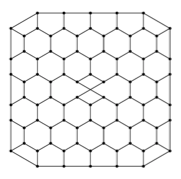}
\includegraphics[width=0.245\textwidth]{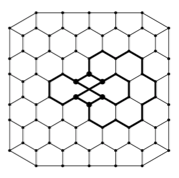}
\caption{On the left is a trivalent spatial graph containing an irreducible crossing of edges. On the right is an illustration of how this irreducible crossing is ultimately due to the presence of a nonplanar ``tangle'' that is a subdivision of ${K_{3, 3}}$. Adapted from S. Wolfram, \textit{A New Kind of Science}, page 527.}
\label{fig:CurrentStatus9}
\end{figure}

\begin{definition}
A ``subdivision'' of an edge ${e \in E}$, where the endpoints of $e$ are given by ${u, v \in V}$, is obtained by introducing a new vertex ${w \in W}$, and replacing $e$ by a new pair of edges ${f_1, f_2 \in F}$, whose endpoints are given by ${u, w \in W}$ and ${w, v \in W}$, respectively.
\end{definition}
This implies that any nonplanarity in a spatial graph must be associated with a finite set of isolable, nonplanar ``tangles'', each of which is a subdivision of either ${K_5}$ or ${K_{3, 3}}$, as shown in Figure \ref{fig:CurrentStatus9}.

Thus, if the update rules for spatial graphs have the property that they always preserve planarity, then this immediately implies that each such nonplanar tangle will effectively act like an independent persistent structure that can only disappear through some kind of ``annihilation'' process with a different nonplanar tangle. As first pointed out in NKS, these persistent nonplanar structures are therefore highly suggestive of elementary particles in particle physics, with the purely graph-theoretic property of planarity playing the role of some conserved physical quantity (such as electric charge). The mathematical details of how one is able to generalize this correspondence, in order to yield a graph-theoretic analog of Noether's theorem (by virtue of Wagner's theorem\cite{wagner}, the Robertson-Seymour theorem\cite{robertson}\cite{robertson2}, and a correspondence between conserved currents in physics and graph families that are closed under the operation of taking graph minors in combinatorics) is outlined in our accompanying publication on quantum mechanics\cite{gorard_new}.

In this context, relativistic mass can be understood in terms of the so-called ``de Bruijn indices''\cite{bruijn} of mathematical logic. In the ${\lambda}$-calculus, abstractions of the form ${\left( \lambda x . M \right)}$ cause the variable $x$ to become bound within the expression, and therefore one must perform ${\alpha}$-conversion of the ${\lambda}$-terms, i.e. one must perform renaming of the bound variables\cite{gabbay}:

\begin{equation}
\left( \lambda x . M \left[ x \right] \right) \to \left( \lambda y . M \left[ y \right] \right),
\end{equation}
in order to avoid name collisions. However, if one instead indexes the bound variables in such a way as to ensure that ${\lambda}$-terms are invariant under ${\alpha}$-conversion, i.e. such that checking for ${\alpha}$-equivalence is the same as checking for syntactic equivalence, then one can avoid having to perform ${\alpha}$-conversion at all. The central idea underlying the de Bruijn index notation is to index each occurrence of a variable within a ${\lambda}$-term by a natural number, designating the number of binders (within scope) that lie between that variable occurrence and its associated binder. For instance, the $K$ and $S$ combinators from combinatory logic, namely:

\begin{equation}
\lambda x . \lambda y . x, \qquad \text{ and } \qquad \lambda x . \lambda y . \lambda z . x z \left( y z \right),
\end{equation}
respectively, can be represented using de Bruijn indices as:

\begin{equation}
\lambda \lambda 2, \qquad \text{ and } \qquad \lambda \lambda \lambda 3 \, 1 \left( 2 \, 1 \right),
\end{equation}
respectively.

De Bruijn indices are correspondingly used within automated theorem-proving systems as a naming convention for generated variables. In constructing an automated proof of an equational theorem, such as ${x = y}$, a standard technique is to treat the problem as one of abstract term rewriting: one can think of the axioms of the system as being abstract rewriting rules, and then the problem of finding a proof of ${x = y}$ is ultimately the problem of determining a rewrite sequence that transforms $x$ to $y$ and vice versa, i.e one wishes to determine that ${x \leftrightarrow^{*} y}$. In constructing this proof, one may consider rewriting rules of the form ${f \left( x, y \right) \to g \left( x, y, z \right)}$, in which the variable $z$ is a generated variable (i.e. it appears on the right-hand-side, but not the left-hand-side, of the rule), which automated theorem-proving systems generally tend to index using de Bruijn index notation. These generated variables will eventually be eliminated, for instance by rules of the form ${g \left( x, y, z \right) \to h \left( x, y \right)}$, but they may be necessary for certain intermediate steps of the proof.

The key point here, however, is that the longer the path that one traverses through the proof graph, the more intermediate (generated) variables will be produced; as such, the generated variables can be thought of as being consequences (artefacts) of particular choices of updating orders. Since all abstract rewriting formalisms are mathematically equivalent (and therefore, in particular, hypergraph substitution systems are ultimately isomorphic to automated theorem-proving systems), this statement corresponds to the claim that, the longer the path that one traverses through a multiway system (i.e. the steeper the angle of ``slicing'' in the foliation of the causal graph), the more intermediate/generated vertices will be produced. As such, elementary particles will appear to be ``heavier'' (i.e. to have more vertices associated with them) on steeper ``slices'' in the foliation of the causal graph, without changing the eventual outcomes of the updating events. This is the discrete analog of relativistic mass increase due to a Lorentz transformation - a concrete demonstration of this idea is given in \cite{wolfram_new}.

\section{Discrete Geometry, Curvature and Gravitation in the Wolfram Model}

\subsection{A Discrete Ricci Scalar Curvature for (Directed) Hypergraphs}
\label{sec:CurrentStatus5}

Intuitively, the notion of curvature in Riemannian geometry is some measure of the deviation of a manifold from being locally Euclidean, or, more formally, a measure of the degree to which the metric connection fails to be exact. For the case of an $n$-dimensional Riemannian manifold, ${\left( \mathcal{M}, g \right)}$, we can make this statement mathematically precise, in the following way: the Ricci scalar curvature (i.e. the simplest curvature invariant of the manifold) is related to the ratio of the volume of a ball of radius ${\epsilon}$, denoted ${B_{\epsilon} \subset \mathcal{M}}$, to the volume of a ball of the same radius in flat (Euclidean) space\cite{ricci}:

\begin{equation}
\frac{\mathrm{Vol} \left( B_{\epsilon} \left( p \right) \subset \mathcal{M} \right)}{\mathrm{Vol} \left( B_{\epsilon} \left( 0 \right) \subset \mathbb{R}^n \right)} = 1 - \frac{R}{6 \left( n + 2 \right)} \epsilon^2 + O \left( \epsilon^4 \right),
\end{equation}
as ${\epsilon \to 0}$, where $R$ denotes the Ricci scalar curvature of ${\left( \mathcal{M}, g \right)}$ at the point ${p \in \mathcal{M}}$. However, an equivalent method of defining $R$, which extends more readily to the discrete case of a directed hypergraph, is related to the ratio of the average distance between points on two balls to the distance between their centres, when those balls are sufficiently small and close together. More specifically, if ${B_{\epsilon} \left( p \right) \subset \mathcal{M}}$ is a ball of radius ${\epsilon}$, centred at point ${p \in \mathcal{M}}$, and it is mapped via parallel transport to ${B_{\epsilon} \left( q \right) \subset \mathcal{M}}$, a corresponding ball of radius ${\epsilon}$ centred at point ${q \in \mathcal{M}}$, then the average distance between a point on ${B_{\epsilon} \left( p \right)}$ and its image (i.e. its corresponding point on ${B_{\epsilon} \left( q \right)}$) is given by:

\begin{equation}
\delta \left( 1 - \frac{\epsilon^2}{2 \left( n + 2 \right)} R + O \left( \epsilon^3 + \epsilon^2 \delta \right) \right),
\end{equation}
as ${\epsilon, \delta \to 0}$, where ${\delta = d \left( p, q \right)}$ denotes the distance between the centres $p$ and $q$.

This local characterization of curvature in terms of average distances between balls, crucially, allows one to extend the notion of Ricci scalar curvature to more general metric spaces (including hypergraphs)\cite{ollivier}\cite{ollivier2}\cite{ollivier3}. If one now considers an arbitrary metric space ${\left( X, d \right)}$, with distance function $d$, then the natural generalization of the concept of a volume measure is a probability measure, with the natural generalization of an average distance between volume measures being the so-called ``Wasserstein distance'' between probability measures:

\begin{definition}
For a Polish metric space ${\left( X, d \right)}$, equipped with its Borel ${\sigma}$-algebra, a ``random walk'' on X, denoted $m$, is a family of probability measures:

\begin{equation}
m = \lbrace m_x : x \in X \rbrace,
\end{equation}
each satisfying the requirement that ${m_x}$ has a finite first moment, and that the map ${x \to m_x}$ is measurable.
\end{definition}

\begin{definition}
The ``1-Wasserstein distance'', denoted ${W_1 \left( m_x, m_y \right)}$, between two probability measures ${m_x}$ and ${m_y}$ on a metric space $X$, is the optimal transportation distance between those measures, given by:

\begin{equation}
W_1 \left( m_x, m_y \right) = \inf_{\epsilon \in \prod \left( m_x, m_y \right)} \left[ \int_{\left( x, y \right) \in X \times X} d \left( x, y \right) d \epsilon \left( x, y \right) \right],
\end{equation}
where ${\prod \left( m_x, m_y \right)}$ denotes the set of measures on ${X \times X}$, i.e. the coupling between random walks projecting to ${m_x}$ and those projecting to ${m_y}$.
\end{definition}
Instinctively, one may view ${\prod \left( m_x, m_y \right)}$ as designating the set of all possible ``transportations'' of the measure ${m_x}$ to the measure ${m_y}$ (where a ``transportation'' involves ``disassembling'' the measure, transporting it, and ``reassembling it'' in the form of ${m_y}$). Thus, the Wasserstein distance is essentially the minimal cost (in terms of transportation distance) required to transport measure ${m_x}$ to measure ${m_y}$.

\begin{definition}
For a metric space ${\left( X, d \right)}$, equipped with a random walk $m$, the ``Ollivier-Ricci scalar curvature'' in the direction ${\left( p, q \right)}$, for distinct points ${p, q \in X}$, is given by:

\begin{equation}
\kappa \left( p, q \right) = 1 - \frac{W_1 \left( m_p, m_q \right)}{d \left( p, q \right)}.
\end{equation}
\end{definition}
In the particular case in which ${\left( X, d \right)}$ is a Riemannian manifold, and $m$ is the standard Riemannian volume measure, the Ollivier-Ricci scalar curvature ${\kappa \left( p, q \right)}$ reduces (up to some arbitrary scaling factor) to the standard Riemannian Ricci scalar curvature ${R \left( p, q \right)}$. As such, one can think of ${m_p}$ and ${m_q}$ as being the appropriate generalization of the notion of volumes of balls centred at points $p$ and $q$ in a Riemannian manifold.

Moreover, for the case in which $X$ is a discrete metric space, one can give the following, more explicit, form of the Wasserstein transportation metric:

\begin{definition}
The ``discrete 1-Wasserstein distance'', denoted ${W_1 \left( m_x, m_y \right)}$, between two discrete probability measures ${m_x}$ and ${m_y}$ on a discrete metric space $X$, is the multi-marginal (i.e. discrete) optimal transportation distance between those measures, given by:

\begin{equation}
W_1 \left( m_x, m_y \right) = \inf_{\mu_{x, y} \in \prod \left( m_x, m_y \right)} \left[ \sum_{\left( x^{\prime}, y^{\prime} \right) \in X \times X} d \left( x^{\prime}, y^{\prime} \right) \mu_{x, y} \left( x^{\prime}, y^{\prime} \right) \right],
\end{equation}
where ${\prod \left( m_x, m_y \right)}$ here denotes the set of all discrete probability measures, ${\mu_{x, y}}$, satisfying:

\begin{equation}
\sum_{y^{\prime} \in X} \mu_{x, y} \left( x^{\prime}, y^{\prime} \right) = m_x \left( x^{\prime} \right),
\end{equation}
and:

\begin{equation}
\sum_{x^{\prime} \in X} \mu_{x, y} \left( x^{\prime}, y^{\prime} \right) = m_y \left( y^{\prime} \right).
\end{equation}
\end{definition}

Let us now demonstrate how this formalism works in practice, by considering the particular case of a directed hypergraph ${H = \left( V, E \right)}$\cite{eidi}, in which every hyperedge ${e \in E}$ designates a directional relation between two sets of vertices, $A$ and $B$, referred to as the ``tail'' and the ``head'' of the hyperedge, respectively. Then, for a given vertex in the tail of a directed hyperedge, ${x_i \in A}$, the number of incoming hyperedges to ${x_i}$, denoted ${d_{x_{i}^{in}}}$, designates the number of hyperedges that include ${x_i}$ as an element of their head set. Similarly, for a given vertex in the head of a directed hyperedge, ${y_i \in B}$, the number of outgoing hyperedges from ${y_j}$, denoted ${d_{y_{j}^{out}}}$, designates the number of hyperedges which include ${y_j}$ as an element of their tail set:

\begin{definition}
For a directed hypergraph ${H = \left( V, E \right)}$, the ``Ollivier-Ricci scalar curvature'' of the directed hyperedge ${e \in E}$, where:

\begin{equation}
A = \lbrace x_1, \dots, x_n \rbrace \to^{e} B = \lbrace y_1, \dots, y_m \rbrace,
\end{equation}
where ${n, m \leq \lvert V \rvert}$, is given by:

\begin{equation}
\kappa \left( e \right) = 1 - W \left( \mu_{A^{in}}, \mu_{B^{out}} \right),
\end{equation}
where the probability measures ${\mu_{A^{in}}}$ and ${\mu_{B^{out}}}$ satisfy:

\begin{equation}
\mu_{A^{in}} = \sum_{i = 1}^{n} \mu_{x_i}, \qquad \text{ and } \qquad \mu_{B^{out}} = \sum_{j = 1}^{m} \mu_{y_j},
\end{equation}
with:

\begin{align}
\forall 1 \leq i \leq n, z \in V, \qquad \mu_{x_i} \left( z \right) = \begin{cases}
0, \qquad & \text{ if } z = x_i \text{ and } d_{x_{i}^{in}} \neq 0,\\
\frac{1}{n}, \qquad & \text{ if } z = x_i \text{ and } d_{x_{i}^{in}} = 0,\\
\sum\limits_{e^{\prime} : z \to x_i} \frac{1}{n \times d_{x_{i}^{in}} \times \lvert tail \left( e^{\prime} \right) \rvert}, \qquad & \text{ if } z \neq x_i \text{ and } \exists e^{\prime} : z \to x_i,\\
0, \qquad & \text{ if } z \neq x_i \text{ and } \nexists e^{\prime} : z \to x_i,
\end{cases}
\end{align}
and:

\begin{align}
\forall 1 \leq j \leq m, z \in V, \qquad \mu_{y_j} \left( z \right) = \begin{cases}
0, \qquad & \text{ if } z = y_j \text{ and } d_{y_{j}^{out}} \neq 0,\\
\frac{1}{m}, \qquad & \text{ if } z = y_j \text{ and } d_{y_{j}^{out}} = 0,\\
\sum\limits_{e^{\prime} : y_j \to z} \frac{1}{m \times d_{y_{j}^{out}} \times \lvert head \left( e^{\prime} \right) \rvert}, \qquad & \text{ if } z \neq y_j \text{ and } \exists e^{\prime} : y_j \to z,\\
0, \qquad & \text{ if } z \neq y_j \text{ and } \nexists e^{\prime} : y_j \to z.
\end{cases}
\end{align}
\end{definition}

\begin{definition}
The ``discrete 1-Wasserstein distance'', denoted ${W_1 \left( \mu_{A^{in}}, \mu_{B^{out}} \right)}$, between two discrete probability measures ${\mu_{A^{in}}}$ and ${\mu_{B^{out}}}$ on a directed hypergraph ${H = \left( V, E \right)}$, is the multi-marginal (i.e. discrete) optimal transportation distance between those measures, given by:

\begin{equation}
W \left( \mu_{A^{in}}, \mu_{B^{out}} \right) = \min \left[ \sum_{u \to A} \sum_{B \to v} d \left( u, v \right) \epsilon \left( u, v \right) \right],
\end{equation}
where ${d \left( u, v \right)}$ represents the minimum number of directed hyperedges that must be traversed when travelling from vertex ${u \in A^{in} \left( u \to A \right)}$ to vertex ${v \in B^{out} \left( B \to v \right)}$, ${\epsilon \left( u, v \right)}$ represents the coupling, i.e. the total mass being moved from vertex $u$ to vertex $v$, and one is minimizing over the set of all couplings, ${\epsilon}$, between measures ${\mu_{A^{in}}}$ and ${\mu_{B^{out}}}$, satisfying:

\begin{equation}
\sum_{u \to A} \epsilon \left( u, v \right) = \sum_{j = 1}^{m} \mu_{y_j} \left( v \right),
\end{equation}
and:

\begin{equation}
\sum_{B \to v} \epsilon \left( u, v \right) = \sum_{i = 1}^{n} \mu_{x_I} \left( u \right).
\end{equation}
\end{definition}

\begin{figure}[ht]
\centering
\includegraphics[width=0.495\textwidth]{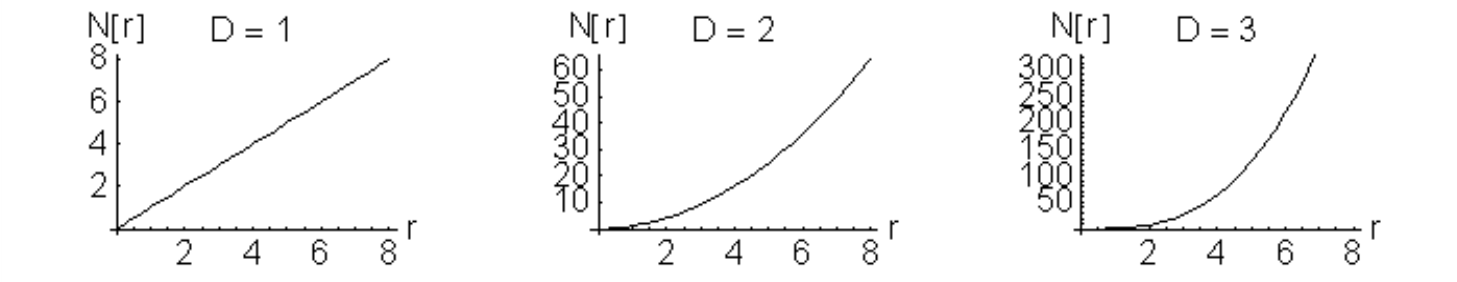}
\caption{The number of vertices that lie within a distance $r$ of a chosen vertex, denoted ${N \left( r \right)}$, plotted for flat spatial hypergraphs in one, two and three dimensions, respectively. Adapted from {\O}. Tafjord, \textit{Fundamental Physics and NKS}, slide 6.}
\label{fig:CurrentStatus10}
\end{figure}

Therefore, using the combinatorial metric ${d \left( u, v \right)}$ defined above, in which each hyperedge is assumed to correspond effectively to a unit of spatial distance, one is able to determine the dimensionality of a given spatial hypergraph by determining the number of vertices, denoted ${N \left( r \right)}$, that lie within a distance $r$ of a chosen vertex. For a ``flat'' spatial hypergraph, ${N \left( r \right)}$ will scale exactly like ${r^n}$:

\begin{equation}
N \left( r \right) = a r^n, \qquad a \in \mathbb{Z},
\end{equation}
where $n$ is the dimensionality of the spatial hypergraph, as shown in Figure \ref{fig:CurrentStatus10}.

\begin{figure}[ht]
\centering
\includegraphics[width=0.495\textwidth]{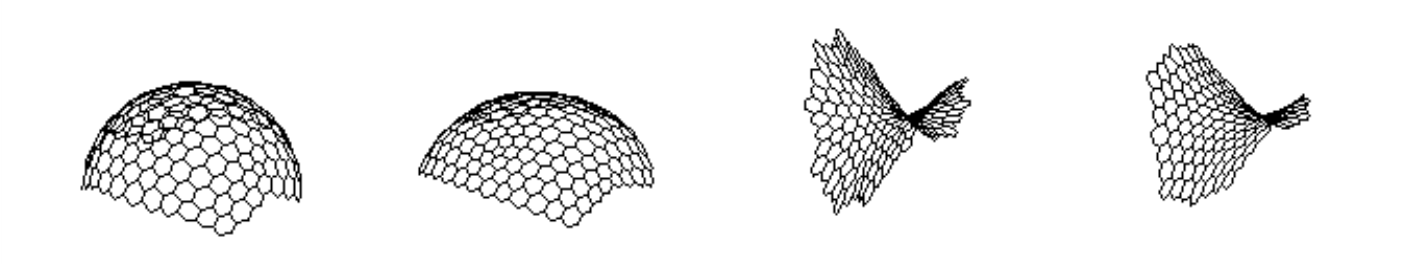}
\caption{Two-dimensional spatial hypergraphs with different Ollivier-Ricci scalar curvatures, with the presence of pentagons indicating regions of positive scalar curvature, and the presence of heptagons indicating regions of negative scalar curvature. Adapted from {\O}. Tafjord, \textit{Fundamental Physics and NKS}, slide 7.}
\label{fig:CurrentStatus11}
\end{figure}

For instance, a two-dimensional spatial hypergraph can be thought of as consisting entirely of hexagons, in which case ${N \left( r \right)}$ will scale asymptotically like the area of a circle, i.e. ${\pi r^2}$. However, a curved spatial hypergraph in two dimensions will also contain pentagonal (corresponding to regions of positive Ollivier-Ricci scalar curvature) and/or heptagonal (corresponding to regions of negative Ollivier-Ricci scalar curvature) structures, wherein there will exist a correction factor to this dimensionality calculation, as shown in Figure \ref{fig:CurrentStatus11}.

Using the fact that the Ollivier-Ricci scalar curvature reduces to the standard Riemannian Ricci scalar curvature in the case of an $n$-dimensional Riemannian manifold, ${\left( \mathcal{M}, g \right)}$, we infer that the ratio of the volume of a discrete ball of radius $r$ in a curved spatial hypergraph to the volume of a discrete ball of the same radius in a flat spatial hypergraph is given by:

\begin{equation}
\frac{\mathrm{Vol} \left( B_r \left( p \right) \in \mathcal{M} \right)}{\mathrm{Vol} \left( B_r \left( 0 \right) \in \mathbb{Z}^n \right)} = 1 - \frac{R}{6 \left( n + 2 \right)} r^2 + O \left( r^4 \right),
\end{equation}
for any sufficiently small value of $r$ (with exactness when ${r = 1}$), thus allowing us to compute the correction factor to ${N \left( r \right)}$ directly as:

\begin{equation}
N \left( r \right) = a r^n \left( 1 - \frac{R}{6 \left( n + 2 \right)} r^2 + O \left( r^4 \right) \right).
\end{equation}
In other words, the correction factor to the dimensionality calculation is found to be proportional to the Ollivier-Ricci scalar curvature, as required. However, in order to derive the complete mathematical formalism of general relativity, one must consider curvature not only in space (as represented by spatial hypergraphs), but in spacetime (as represented by causal graphs). This, in turn, requires introducing appropriately discretized notions of parallel transport, holonomy, the metric tensor, the Riemann curvature tensor, and the spacetime Ricci curvature tensor, which we proceed to do below.

\subsection{Parallel Transport, Discrete Holonomy, and the Discrete Spacetime Ricci Tensor}
\label{sec:CurrentStatus6}

In standard differential geometry, the Ricci scalar curvature, $R$, is often introduced as being the trace of the Ricci curvature tensor, denoted ${R_{a b}}$:

\begin{equation}
R = \mathrm{Tr} \left( R_{a b} \right) = g^{a b} R_{a b} = R_{a}^{a},
\end{equation}
for a metric tensor ${g_{a b}}$. In much the same way as the Ricci scalar curvature can be interpreted as a measure of the ratio of the volume of a ball in an $n$-dimensional Riemannian manifold, ${\left( \mathcal{M}, g \right)}$ to the volume of a ball of the same radius in flat (Euclidean) space, the Ricci curvature tensor also has the following direct geometrical interpretation: ${R \left( \boldsymbol\xi, \boldsymbol\xi \right)}$ is a measure of the ratio of the volume of a conical region in the direction of the vector ${\boldsymbol\xi}$, consisting of geodesic segments of length ${\epsilon}$ emanating from a single point ${p \in \mathcal{M}}$, to the volume of the corresponding conical region in flat (Euclidean) space.

The Ricci curvature may, in turn, be defined as being the contraction of the Riemann curvature tensor, denoted ${R_{a b c d}}$, between the first and third indices:

\begin{equation}
R_{a b} = R_{a c b}^{c},
\end{equation}
where the Riemann curvature tensor effectively quantifies how the direction of an arbitrary vector, ${\mathbf{v}}$, changes as it gets parallel transported around a closed curve in ${\left( \mathcal{M}, g \right)}$:

\begin{equation}
\delta v^{\alpha} = R_{\beta \gamma \delta}^{\alpha} d x^{\gamma} d x^{\delta} v^{\beta};
\end{equation}
in other words, the Riemann curvature tensor is a measure of the degree to which a Riemannian manifold fails to be ``holonomic'' (i.e. fails to preserve geometrical data, as that data gets transported around a closed curve). A more algebraic way of stating the same thing is that the Riemann curvature tensor is the commutator of the covariant derivative operator acting on an arbitrary vector:

\begin{equation}
\nabla_c \nabla_d v^a - \nabla_d \nabla_c v^a = R_{b c d}^{a} v^b.
\end{equation}
When defining the Ricci curvature tensor, the reason for contracting ${R_{a b c d}}$ between the first and third indices specifically is that, due to the intrinsic symmetries of the Riemann curvature tensor, it is not difficult to prove that any other choice of contraction will either yield zero, or ${\pm R_{a b}}$.

In order to have some coherent notion of parallel transport on a standard Riemannian manifold (and hence to introduce such concepts as holonomy and Riemann curvature), it is necessary first to define a connection on that manifold. The standard choice in Riemannian geometry is the Levi-Civita connection (otherwise known as the torsion-free or metric connection), whose components in a chosen basis are given by the Christoffel symbols, denoted ${\Gamma_{\mu \nu}^{\rho}}$, and which can be expressed in terms of the first derivatives of the metric tensor as:

\begin{equation}
\Gamma_{\mu \nu}^{\rho} = \frac{1}{2} g^{\rho \sigma} \left( \partial_{\mu} g_{\sigma \nu} + \partial_{\nu} g_{\mu \sigma} - \partial_{\sigma} g_{\mu \nu} \right).
\end{equation}
Intuitively, the Christoffel symbols designate how the basis vectors change as one moves around on the manifold. In the Levi-Civita connection, the Riemann curvature tensor can be written explicitly in terms of second derivatives the metric tensor as:

\begin{equation}
R_{b c d}^{a} = \partial_c \Gamma_{b d}^{a} - \partial_d \Gamma_{b c}^{a} + \Gamma_{e c}^{a} \Gamma_{b d}^{e} - \Gamma_{e d}^{a} \Gamma_{b c}^{e}.
\end{equation}
In a locally flat, inertial reference frame (i.e. one in which the Christoffel symbols, but not their derivatives, all vanish), these components can be expressed in lowered-index form as:

\begin{equation}
R_{\alpha \beta \gamma \delta} = \frac{1}{2} \left( \partial_{\beta} \partial_{\gamma} g_{\alpha \delta} - \partial_{\beta} \partial_{\delta} g_{\alpha \gamma} + \partial_{\alpha} \partial_{\delta} g_{\beta \gamma} - \partial_{\alpha} \partial_{\gamma} g_{\beta \delta} \right).
\end{equation}

These concepts now allow us to make the geometrical interpretation of the Ricci curvature tensor more precise: if one chooses to use geodesic normal coordinates around the point ${p \in \mathcal{M}}$, i.e. coordinates in which all geodesics passing through $p$ correspond to straight lines passing through the origin, then the metric tensor can be approximated by the Euclidean metric:

\begin{equation}
g_{i j} = \delta_{i j} + O \left( \lVert \mathbf{x} \rVert^2 \right),
\end{equation}
where ${\delta_{i j}}$ denotes the standard Kronecker delta function, indicating that the geodesic distance from $p$ can be approximated by the standard Euclidean distance, as one would expect. The correction factor can be determined explicitly by performing a Taylor expansion of the metric tensor along a radial geodesic in geodesic normal coordinates:

\begin{equation}
g_{i j} = \delta_{i j} - \frac{1}{3} R_{i j k l} x^k x^l + O \left( \lVert \mathbf{x} \rVert^3 \right),
\end{equation}
where the Taylor expansion is performed in terms of Jacobi fields:

\begin{equation}
\mathbf{J} \left( t \right) = \left( \frac{\partial \boldsymbol\gamma_{\tau} \left( t \right)}{\partial \tau} \right)_{\tau = 0},
\end{equation}
i.e. in terms of the tangent space to a given geodesic, denoted ${\boldsymbol\gamma_0}$, in the space of all possible geodesics (here, ${\boldsymbol\gamma_{\tau}}$ denotes a one-parameter family of geodesics). Expanding the square root of the determinant of the metric tensor hence yields an expansion of the metric volume element of ${\left( \mathcal{M}, g \right)}$, denoted ${d \mu_g}$, in terms of the volume element of the Euclidean metric, denoted ${d \mu_{Euclidean}}$:

\begin{equation}
d \mu_g = \left[ 1 - \frac{1}{6} R_{j k} x^j x^k + O \left( \lVert \mathbf{x} \rVert^3 \right) \right] d \mu_{Euclidean},
\end{equation}
as required.

Therefore, in order to be able to proceed sensibly with the derivation, we must now introduce appropriately discretized notions of holonomy, Riemann curvature, etc., that maintain compatibility with the definition of the Ollivier-Ricci scalar curvature outlined previously. Note that, in the formal definition of the Ollivier-Ricci scalar curvature for (directed) hypergraphs given above, we have implicitly employed a discrete notion of parallel transport when mapping points on one hypergraph ball onto corresponding points on a nearby hypergraph ball. To make this definition more explicit, and hence to derive the discrete analog of the full Riemann curvature tensor, it is helpful first to consider the notion of sectional curvature on Riemannian manifolds:

\begin{equation}
K \left( \mathbf{u}, \mathbf{v} \right) = \frac{\langle R \left( \mathbf{u}, \mathbf{v} \right) \mathbf{v}, \mathbf{u} \rangle}{\langle \mathbf{u}, \mathbf{u} \rangle \langle \mathbf{v}, \mathbf{v} \rangle - \langle \mathbf{u}, \mathbf{v} \rangle^2},
\end{equation}
where $R$ is the full Riemann curvature tensor, and ${\mathbf{u}}$ and ${\mathbf{v}}$ are linearly independent tangent vectors at some point ${p \in \mathcal{M}}$. Geometrically, the sectional curvature at point ${p \in \mathcal{M}}$ designates the Gaussian curvature of the surface obtained from the set of all geodesics starting at $p$, and proceeding in the directions of the tangent plane ${\sigma_p}$ defined by the tangent vectors ${\mathbf{u}}$ and ${\mathbf{v}}$. In particular, when ${\mathbf{u}}$ and ${\mathbf{v}}$ are orthonormal, one has:

\begin{equation}
K \left( \mathbf{u}, \mathbf{v} \right) = \langle R \left( \mathbf{u}, \mathbf{v} \right) \mathbf{v}, \mathbf{u} \rangle.
\end{equation}

Consider now an arbitrary metric space ${\left( X, d \right)}$. If ${\boldsymbol\gamma}$ denotes a unit-speed geodesic whose origin is the point ${x \in X}$, and whose initial direction is ${\mathbf{v}}$, then we denote the endpoint of that geodesic by ${exp_x \left( \mathbf{v} \right)}$, so-called because of its relation to the exponential map from tangent spaces to manifolds in Riemannian geometry. Now, a natural generalization of the Riemannian sectional curvature, ${K \left( \mathbf{u}, \mathbf{v} \right)}$, to the metric space ${\left( X, d \right)}$ is as follows:

\begin{equation}
d \left( exp_x \left( \epsilon \mathbf{w}_x \right), exp_y \left( \epsilon \mathbf{w}_y \right) \right) = \delta \left( 1 - \frac{\epsilon^2}{2} K \left( \mathbf{v}, \mathbf{w} \right) + O \left( \epsilon^3 + \epsilon^2 \delta \right) \right),
\end{equation}
as ${\epsilon, \delta \to 0}$. In the above, ${\mathbf{v}}$ and ${\mathbf{w}_x}$ denote unit-length tangent vectors at some point ${x \in X}$, point ${y \in X}$ denotes the endpoint of the vector ${\delta \mathbf{v}}$, ${\mathbf{w}_y}$ is the vector obtained by the parallel transport of vector ${\mathbf{w}_x}$ from point $x$ to point $y$, and ${\delta = d \left( x, y \right)}$ denotes the distance between the points $x$ and $y$. In this more general context, the sectional curvature is a measure of the discrepancy between the distance between points $x$ and $y$, and the distance between the points lying a distance ${\epsilon}$ away from $x$ and $y$ along the geodesics starting at ${\mathbf{w}_x}$ and ${\mathbf{w}_y}$, respectively. The Ollivier-Ricci scalar curvature for an arbitrary metric space can then be recovered by taking the average of the sectional curvature ${K \left( \mathbf{v}, \mathbf{w} \right)}$ over all vectors ${\mathbf{w}}$. More formally, if ${S_x}$ is the set of all tangent vectors of length ${\epsilon}$ at point ${x \in X}$, and similarly for ${S_y}$, and if ${S_x}$ is mapped to ${S_y}$ via parallel transport, then the average distance between a point in ${S_x}$ and its image is given by:

\begin{equation}
\delta \left( 1 - \frac{\epsilon^2}{2 \left( n + 2 \right)} R + O \left( \epsilon^3 + \epsilon^2 \delta \right) \right),
\end{equation}
as ${\epsilon, \delta \to 0}$, where $R$ is the Ollivier-Ricci scalar curvature, as required.

In the case of a (directed) hypergraph, one can set, as usual, all hyperedges to be of unit length, and hence one can also take ${\epsilon = \delta = 1}$ in all of the definitions given above. Geodesics in the hypergraph are then given by solutions to the shortest paths problem. We also assume that all pairs of hyperedges which are not parallel (i.e. incident to the same set of vertices) can be considered to be orthonormal. These assumptions allow us to translate the generalized definition of sectional curvature given above to the case of (directed) hypergraphs, in a manner that is fully consistent with the discrete definition of the Ollivier-Ricci scalar curvature as discussed earlier. Rather gratifyingly, since the components of the sectional curvature tensor completely determine the components of the Riemann curvature tensor:

\begin{equation}
K \left( \mathbf{u}, \mathbf{v} \right) = \langle R \left( \mathbf{u}, \mathbf{v} \right) \mathbf{v}, \mathbf{u} \rangle,
\end{equation}
one immediately obtains discrete analogs of the Riemann curvature and Ricci curvature tensors for the case of (directed) hypergraphs.

Finally, in addition to asking about how the number of vertices, ${N \left( r \right)}$ contained within a ball of radius $r$ in a particular spatial hypergraph grows as a function of $r$, we now also have the requisite technical machinery to ask about how the number of updating events, denoted ${C \left( t \right)}$, contained within a cone of length $t$ within a particular causal graph grows as a function of $t$ (where $t$ can be thought of as corresponding to the number of discrete spacelike hypersurfaces intersected by the cone, for a particular discrete foliation of the casual graph). Henceforth, we assume that special relativity holds (i.e. that all update rules referenced are causal-invariant).

\begin{figure}[ht]
\centering
\includegraphics[width=0.295\textwidth]{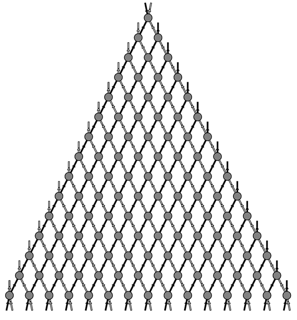}
\caption{A cone of updating events, embedded within a flat, two-dimensional causal graph. Adapted from S. Wolfram, \textit{A New Kind of Science}, page 518.}
\label{fig:CurrentStatus12}
\end{figure}

By direct analogy to the spatial hypergraph case, for a causal graph corresponding to flat, $n$-dimensional spacetime, the number of updating events reached will clearly scale like ${t^n}$:

\begin{equation}
C \left( t \right) = a t^n, \qquad a \in \mathbb{Z},
\end{equation}
as shown in Figure \ref{fig:CurrentStatus12}.

However, in the presence of non-zero spacetime curvature, there exists (again, by analogy to the purely spatial case) a correction factor, in the form of a coefficient of ${t^{n + 2}}$, which we now proceed to compute explicitly, by exploiting the fact that the discrete spacetime Ollivier-Ricci curvature tensor, ${R_{i j}}$, determines the ratio of the volume of a conical region in the causal graph consisting of geodesic segments of unit length emanating from a single vertex, to the volume of the corresponding conical region in a purely flat causal graph. Since, in the Riemannian case, one has:

\begin{equation}
d \mu_g = \left[ 1 - \frac{1}{6} R_{j k} x^j x^k + O \left( \lVert \mathbf{x} \rVert^3 \right) \right] d \mu_{Euclidean},
\end{equation}
one now obtains, in the discrete case:

\begin{equation}
C \left( t \right) = a t^n \left[ 1 - \frac{1}{6} R_{j k} t^j t^k + O \left( \lVert \mathbf{t} \rVert^3 \right) \right],
\end{equation}
i.e. the correction factor is indeed found to be a coefficient of ${t^{n + 2}}$, and is proportional to the projection of the discrete spacetime Ollivier-Ricci curvature tensor along the discrete time vector, ${\mathbf{t}}$, defined by the particular choice of foliation of the causal graph into discrete spacelike hypersurfaces. In terms of the discrete ADM decomposition introduced earlier, this time vector has the generic form:

\begin{equation}
t^a = \alpha n^a + \beta^a,
\end{equation}
for a given choice of lapse function ${\alpha}$ and shift vector ${\beta^i}$.

All that remains is to deduce an appropriate set of constraints on the spacetime Ollivier-Ricci curvature tensor, and to show (subject to certain assumptions about the limiting conditions of the causal graph) that these constraints are equivalent to an appropriately discretized form of the Einstein field equations.

\subsection{The Discrete Einstein Field Equations}
\label{sec:CurrentStatus7}

In all that follows, we shall assume one further condition on the hypergraph update rules, beyond mere causal invariance: namely, ``asymptotic dimensionality preservation''. Loosely speaking, this means that the dimensionality of the causal graph show converge to some fixed, finite value as the number of updating events grows arbitrarily large. We can express this requirement more formally by stating that the growth rate of the ``global dimension anomaly'', i.e. the correction factor to the global dimensionality of the causal graph, should converge to zero as the size of the causal graph increases (since converging to anything non-zero would imply a global dimension anomaly that grows without bound, and would therefore correspond to a causal graph with effectively unbounded dimensionality). Since, as established above, the local correction factor to the dimensionality of the causal graph is proportional to a projection of the discrete spacetime Ollivier-Ricci curvature tensor ${R_{i j}}$, we begin by computing an average of this quantity across both all possible projection directions and all possible vertices, in order to obtain a value for the global dimension anomaly; averaging out over all timelike projection directions ${\mathbf{t}}$ corresponds to a standard tensor index contraction, thus reducing ${R_{i j}}$ to the discrete spacetime Ollivier-Ricci scalar curvature $R$, and hence yielding a volume average (i.e. global dimension anomaly) given by:

\begin{equation}
S \left[ g^{a b} \right] = \sum d \mu_g R.
\end{equation}
Our condition that the rules be asymptotically dimensionality preserving therefore corresponds to the statement that the change in the value of the global dimension anomaly, with respect to the discrete causal graph metric tensor, should converge to zero in the limit of a large causal graph:
\begin{equation}
\frac{\delta S \left[ g^{a b} \right]}{\delta g^{a b}} \to 0,
\end{equation}

In the continuum limit of an arbitrarily large causal graph, this sum becomes (subject to a weak ergodicity assumption on the dynamics of the causal graph) an integral, with the volume element now being given by the determinant of the metric tensor, as demonstrated earlier:

\begin{equation}
S \left[ g^{a b} \right] = \int d^4 x \sqrt{- g} R,
\end{equation}
such that the constraint that the update rules must be asymptotically dimensionality preserving becomes mathematically equivalent to the statement that the classical (vacuum) Einstein-Hilbert action\cite{hilbert}, with the standard general relativistic Lagrangian density:

\begin{equation}
\mathcal{L}_G = \sqrt{- g} R,
\end{equation}
must be extremized. Therefore, we can adopt the standard approach of taking a functional derivative of the Einstein-Hilbert action with respect to the inverse metric tensor, and enforcing the assumption of zero surface terms, to obtain:

\begin{equation}
\frac{\delta S \left[ g^{ a b} \right]}{\delta g^{a b}} = \sqrt{- g} \left( R_{a b} - \frac{1}{2} R g_{a b} \right),
\end{equation}
with minimization of the action hence yielding the vacuum Einstein field equations:

\begin{equation}
R_{a b} - \frac{1}{2} R g_{a b} = 0.
\end{equation}
Thus, for any set of hypergraph updating rules that are both causal-invariant and asymptotically dimensionality preserving, then if a continuum limit exists in which the causal graph becomes a Riemannian manifold, that limiting manifold must satisfy the vacuum Einstein field equations, as required. A more intuitive statement of this derivation, in which the argument is made that any local appearance of curvature in the causal graph must disappear globally in order to allow the curvature to be ``distinguishable'' from a global dimension change, is given in \cite{wolfram_new}.

\begin{figure}[ht]
\centering
\includegraphics[width=0.495\textwidth]{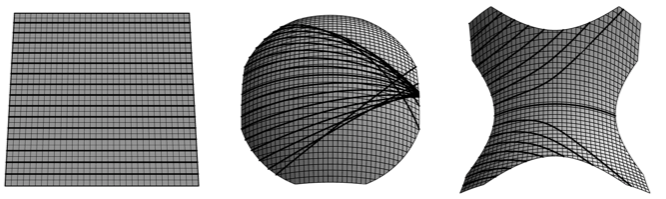}
\caption{Geodesic trajectories of test particles (starting on initially parallel paths) through flat, positively-curved and negatively-curved discrete spaces, respectively. Adapted from S. Wolfram, \textit{A New Kind of Science}, page 531.}
\label{fig:CurrentStatus13}
\end{figure}

Amongst many other consequences, this implies that the trajectory, denoted ${\mathbf{x} \left( s \right)}$, of a test particle (e.g. a persistent nonplanar tangle embedded within a spatial hypergraph) through a causal graph, will, in the continuum limit and in the absence of any non-gravitational forces, obey the relativistic geodesic equation:

\begin{equation}
\frac{d^2 x^{\mu}}{d s^2} + \Gamma_{\alpha \beta}^{\mu} \frac{d x^{\alpha}}{d s} \frac{d x^{\beta}}{d s} = 0.
\end{equation}
This fact can be proved in the usual way, by considering, for instance, a geodesic between a pair of timelike-separated events in the causal graph. One starts from the action for a timelike curve:

\begin{equation}
S = \int d s,
\end{equation}
where ${d s}$ designates the spacetime line element:

\begin{equation}
ds = \sqrt{- g_{\mu \nu} d x^{\mu} d x^{\nu}},
\end{equation}
with the negative sign before the metric tensor indicating the timelike nature of the curve. Then, by introducing a scalar variable, ${\lambda}$, by which to parameterize the action:

\begin{equation}
S = \int \sqrt{- g_{\mu \nu} \frac{d x^{\mu}}{d \lambda} \frac{d x^{\nu}}{d \lambda}} d \lambda,
\end{equation}
we are now able to vary the action with respect to the curve ${x^{\mu}}$, thus yielding the geodesic equation via the principle of least action. Analogous derivations exist for geodesics between lightlike- and spacelike-separated events too. The qualitative consequences of the geodesic equation for spatial hypergraphs are illustrated in Figure \ref{fig:CurrentStatus13}.

\begin{figure}[ht]
\centering
\includegraphics[width=0.495\textwidth]{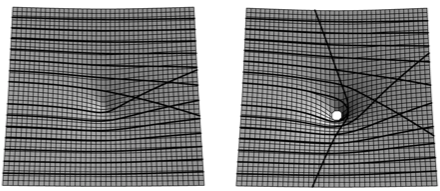}
\caption{Geodesic trajectories of test particles (starting on initially parallel paths) through discrete spaces containing a spherical mass distribution. Adapted from S. Wolfram, \textit{A New Kind of Science}, page 531.}
\label{fig:CurrentStatus14}
\end{figure}

To incorporate matter contributions within the field equations, one can use the first-principles definition of the energy-momentum tensor, i.e. that ${T^{\mu \nu}}$ designates the flux of the relativistic 4-momentum ${P^{\mu}}$ across the hypersurface of constant ${x^{\nu}}$. This definition readily extends to the discrete case, since (as demonstrated in the special relativity derivation above) this flux of the relativistic 4-momentum is, in the simplest case, the number of crossing edges, i.e. the number of edges bound within nonplanar tangles of some spatial hypergraph, which pass through a given discrete hypersurface in the causal graph (a more formal statement of this idea is presented towards the end of this subsection). This increase in the effective number of hyperedges increases the number of reachable vertices, ${N \left( r \right)}$, lying within a distance $r$ of a given vertex, and hence must be corrected for within the dimensionality calculation. We can accommodate this correction term by effectively adding a matter contribution to the relativistic Lagrangian density:

\begin{equation}
\mathcal{L} = \mathcal{L}_G + C_M \mathcal{L}_M,
\end{equation}
where ${C_M}$ is an arbitrary constant (which we assume, by convention, to be equal to ${16 \pi}$ in the case of scalar field matter). Now, minimization of the combined action yields to the full (non-vacuum) Einstein field equations:

\begin{equation}
R_{a b} - \frac{1}{2} R g_{a b} = 8 \pi T_{a b},
\end{equation}
so long as the energy-momentum tensor is taken to be of the following general form in terms of the matter Lagrangian density:

\begin{equation}
T_{a b} = - \frac{C_M}{8 \pi \sqrt{- g}} \frac{\partial \mathcal{L}_M \left[ g^{a b} \right]}{\partial g^{a b}},
\end{equation}
which can also be reformulated in terms of an effective matter action, as:

\begin{equation}
T_{a b} = - \frac{C_M}{8 \pi \sqrt{- g}} \frac{\delta S_M \left[ g^{a b} \right]}{\delta g^{a b}}.
\end{equation}
The qualitative consequences of the inclusion of matter fields are illustrated in Figure \ref{fig:CurrentStatus14}.

Note that this entire derivation defines the Einstein field equations only up to an integration constant; more specifically, the constraint that the update rules should be asymptotically dimensionality preserving could equally well have been enforced by using the sum over the dimensionality correction factor, plus some arbitrary constant, which we (suggestively) denote ${2 \Lambda}$:

\begin{equation}
S \left[ g^{a b} \right] = \sum d \mu_g \left( R + 2 \Lambda \right).
\end{equation}
Running through the same logic as above, we see that, in the continuum limit, the general relativistic Lagrangian density will now become:

\begin{equation}
\mathcal{L}_G = \sqrt{- g} \left( R + 2 \Lambda \right),
\end{equation}
thus yielding the new field equations:

\begin{equation}
R_{a b} - \frac{1}{2} R g_{a b} + \Lambda g_{a b} = 8 \pi T_{a b}.
\end{equation}
Therefore, we can conclude that the Wolfram Model is consistent with both zero and non-zero values of the cosmological constant. This suggests a generalization of the definition of the discrete energy-momentum tensor given above, as being instead a measure of the flux (i.e. intersection) of causal edges through a discrete hypersurface in the causal graph, with causal edges associated with the evolution of elementary particles (i.e. nonplanar tangles) corresponding to standard baryonic matter contributions, and causal edges associated with the evolution of the background space corresponding to vacuum energy contributions. In the usual way, pure energy and momentum may be extracted from ${T^{\mu \nu}}$ by considering fluxes through purely spacelike and purely timelike hypersurfaces, respectively.

It is interesting to note that the nature of this derivation of the continuum Einstein field equations from the underlying discrete geometry of the causal graph is formally analogous to the so-called ``Chapman-Enskog'' hydrodynamic expansion\cite{chapman}\cite{balescu}, used in the derivation of the continuum Navier-Stokes equations from underlying discrete molecular dynamics. Generalizations of Chapman-Enskog theory were previously developed by Wolfram\cite{wolfram7} in the derivation of continuum hydrodynamics equations from discrete cellular automaton models. In standard Chapman-Enskog theory, one starts from the 1-particle distribution function, ${f \left( \mathbf{r}, \mathbf{v}, t \right)}$, and constructs its Boltzmann equation\cite{cercignani}:

\begin{equation}
\frac{\partial f}{\partial t} + \mathbf{v} \cdot \frac{\partial f}{\partial \mathbf{r}} + \mathbf{F} \cdot \frac{\partial f}{\partial \mathbf{v}} = \hat{C} f,
\end{equation}
for some nonlinear integral operator, ${\hat{C}}$, characterizing the collisions between particles and the resultant effect on the evolution of the distribution function. One now proceeds to introduce a dummy parameter, denoted ${\epsilon}$, into the Boltzmann equation:

\begin{equation}
\frac{\partial f}{\partial t} + \mathbf{v} \cdot \frac{\partial f}{\partial \mathbf{r}} + \mathbf{F} \cdot \frac{\partial f}{\partial \mathbf{v}} = \frac{1}{\epsilon} \hat{C} f,
\end{equation}
and then to expand $f$ as a formal power series in ${\epsilon}$:

\begin{equation}
f = f^{(0)} + \epsilon f^{(1)} + \epsilon^2 f^{(2)} + \dots.
\end{equation}
The zeroth-order solution, given by:

\begin{equation}
f^{(0)} = n^{\prime} \left( \mathbf{r}, t \right) \left( \frac{m}{2 \pi k_B T^{\prime} \left( \mathbf{r}, t \right)} \right)^{\frac{3}{2}} \exp \left[ - \frac{m \left( \mathbf{v} - \mathbf{v}_0^{\prime} \left( \mathbf{r}, t \right) \right)^2 }{2 k_B T^{\prime} \left( \mathbf{r}, t \right)} \right],
\end{equation}
yields the dissipation-free Euler equations (i.e. the hydrodynamic description of a gas in equilibrium, in the absence of thermal conductivity and viscosity), for functions ${n^{\prime} \left( \mathbf{r}, t \right)}$, ${\mathbf{v}_{0}^{\prime} \left( \mathbf{r}, t \right)}$ and ${T^{\prime} \left( \mathbf{r}, t \right)}$ determined by the moments of ${f \left( \mathbf{r}, \mathbf{v}, t \right)}$. The second-order solution, given by:

\begin{equation}
f^{(1)} = \left[ - \frac{1}{n} \left( \frac{2 k_B T}{m} \right)^{\frac{1}{2}} \mathbf{A} \left( \mathbf{v} \right) \cdot \nabla \log \left( T \right) - \frac{2}{n} \mathbb{B} \left( \mathbf{v} \right) : \nabla \mathbf{v}_0 \right] f^{(0)},
\end{equation}
yields the dissipative full Navier-Stokes equations, where ${\mathbf{A} \left( \mathbf{v} \right)}$ and ${\mathbb{B} \left( \mathbf{v} \right)}$ are a vector and tensor, respectively, determined by solutions to an integral equation, and ${\cdot : \cdot}$ denotes the double dot product between tensors:

\begin{equation}
\mathbb{T} : \mathbb{T^{\prime}} = \sum_i \sum_j T_{i j} T_{j i}^{\prime}.
\end{equation}
Higher-order terms in the expansion give rise to higher-order supersets of the Navier-Stokes equations, such as the Burnett equations and the super-Burnett equations.

As such, one can now see that our derivation of the Einstein field equations in the continuum limit of causal-invariant and asymptotically dimensionality preserving Wolfram Model systems is essentially equivalent to a Chapman-Enskog hydrodynamic expansion, but with the function:

\begin{equation}
C \left( t \right) = a t^n \left[ 1 - \frac{1}{6} R_{j k} t^j t^k + O \left( \lVert \mathbf{t} \rVert^3 \right) \right],
\end{equation}
playing the role of a ``distribution function'' for vertices in the causal graph. One then recovers the dynamics of the Ricci curvature tensor (standard general relativity) as the zeroth-order term in the expansion - analogous to the Euler equations - and the dynamics of the Weyl curvature tensor (higher-order general relativity) as the corresponding first-order term - analogous to the Navier-Stokes equations. Excitingly, this approach leaves open the possibility of the existence of additional higher-order supersets of the Einstein field equations, presumably analogous to the Burnett and super-Burnett equations, which have been conjectured by Wolfram to have possible implications for dark matter\cite{wolfram6}, for instance by analogy to Milgrom's ``Modified Newtonian Dynamics'' (MOND) proposal\cite{milgrom}.

We intend to compute and investigate these potential higher-order corrections to the discrete Einstein field equations more fully within a future publication, but one possible way in which they may manifest is in the form of higher-order contractions of the Riemann curvature tensor, as in the Lovelock theory of higher-dimensional gravity\cite{lovelock}\cite{woolliams}:

\begin{align}
\alpha \left( - \frac{1}{2} R_{\rho \sigma} R^{\rho \sigma} g_{\mu nu} - \nabla_{\nu} \nabla_{\mu} R - 2 R_{\rho \nu \mu \sigma} R^{\sigma \rho} + \frac{1}{2} g_{\mu \nu} \square R \square R_{\mu \nu} \right) &+\\
 \beta \left( \frac{1}{2} R^2 g_{\mu \nu} - 2 R R_{\mu \nu} - 2 \nabla_{\nu} \nabla_{\mu} R + 2 g_{\mu \nu} \square R \right) &+\\
 \gamma \left( - \frac{1}{2}{\kappa^{-2}} R g_{\mu \nu} + \kappa^{-2} R_{\mu \nu} \right) &= 0,
\end{align}
with free parameters ${\alpha}$, ${\beta}$, ${\gamma}$, and where:

\begin{equation}
\kappa^2 = 32 \pi G.
\end{equation}
A directly geometrical intuition for a possible manifestation of higher-order corrections as follows: the Einstein field equations provide constraints on the Ricci curvature tensor, and hence on the volumes of geodesic bundles in spacetime, but do not provide any explicit constraints on the shapes of those bundles, which are instead described by the Weyl curvature tensor\cite{hawking}\cite{danehkar}:

\begin{align}
C_{i k l m} = R_{i k l m} &+ \frac{1}{n - 2} \left( R_{i m} g_{k l} - R_{i l} g_{k m} + R_{k l} g_{i m} - R_{k m} g_{i l} \right)\\
&+ \frac{1}{\left( n - 1 \right) \left( n - 2 \right)} R \left( g_{i l} g_{k m} - g_{i m} g_{k l} \right).
\end{align}
Consider now the hydrodynamics analogy. In fluid mechanics, the total viscous stress tensor (i.e. the tensor describing the stresses within the fluid due purely to its strain rate) can be exactly decomposed into a sum of a trace part (i.e. a scalar multiple of the identity tensor, manifesting as a bulk hydrostatic pressure), and a trace-free part (i.e. the viscous shear stress tensor). The Euler equations place constraints on the trace part (the pressure), and their higher-order corrections place constraints on the trace-free part (the shear stresses). In exactly the same way, the Riemann curvature tensor in general relativity can be decomposed into a sum of a trace part (i.e. the Ricci curvature tensor, a sum of Kronecker delta functions) and a trace-free part (i.e. the Weyl curvature tensor). The Einstein field equations place constraints on the Ricci curvature tensor, and their higher-order corrections may very well place explicit additional constraints on the Weyl curvature tensor too.

\subsection{Topology, Variable Dimensionality and Implications for Cosmology}
\label{sec:CurrentStatus8}

One assumption that has been implicit in the reasoning presented above is that, when computing the dimensionality of spatial hypergraphs:

\begin{equation}
N \left( r \right) = a r^n \left( 1 - \frac{R}{6 \left( n + 2 \right)} r^2 + O \left( r^4 \right) \right),
\end{equation}
or causal graphs:

\begin{equation}
C \left( t \right) = a t ^n \left[ 1 - \frac{1}{6} R_{j k} t^j t^k + O \left( \lVert \mathbf{t} \rVert^3 \right) \right],
\end{equation}
we have assumed that the dimensionality, $n$, is constant, and that the curvature, $R$, varies from point to point on the manifold. However, it is equally mathematically consistent to fix the value of $R$, and to allow $n$ to vary as a function of position, effectively treating curvature as a highly localized change in the dimensionality of the manifold. Of course, this involves slightly generalizing the notion of a manifold, since for an ordinary manifold ${\mathcal{M}}$, dimensionality is a local invariant. This indicates that the function ${f : \mathcal{M} \to \mathbb{R}}$, mapping every point in ${\mathcal{M}}$ to the dimensionality of its neighborhood, is locally constant, in the sense that\cite{lee}:

\begin{equation}
\forall x \in \mathcal{M}, \exists \text{ neighborhood } U \subset \mathcal{M}, \text{ such that } f \text{ is constant on } U,
\end{equation}
which, in turn, entails that every connected component of ${\mathcal{M}}$ must have a fixed dimensionality.

Developing an extension of general relativity that can accommodate spacetime geodesics with varying fractal dimension has been investigated in the context of so-called ``Scale Relativity''\cite{nottale}\cite{ord}\cite{nottale2}, in which the principle of general covariance is extended to accommodate scale invariance, i.e. in which physical laws are taken to be valid in all coordinate systems, irrespective of their absolute scale. Within general scale relativity, this variation in fractal dimension is constrained by a logarithmic version of the Lorentz transformation, and is treated as a form of ``curvature in scale space''. However, a complete general theory of scale relativity, including scale-invariant versions of the Einstein field equations, has yet to be constructed.

For our present purposes, we shall consider one possible method of formalizing the treatment of curvature as a local change in dimensionality: a unification of the topological concepts of coordinate charts and covering maps. In standard topology, if ${\mathcal{M}}$ is a topological manifold, then a coordinate chart is a homeomorphism\cite{jost}\cite{lee2}:

\begin{equation}
\varphi : \left( U \subset \mathcal{M} \right) \to \left( \varphi \left( U \right) \subset \mathbb{R}^n \right),
\end{equation}
i.e. a homeomorphism from an open subset of ${\mathcal{M}}$ to an open subset of the flat (Euclidean) space of dimension $n$. An atlas of ${\mathcal{M}}$ is then an indexed family of charts:

\begin{equation}
A = \lbrace \left( U_{\alpha}, \varphi_{\alpha} \right) : \alpha \in I \rbrace,
\end{equation}
which covers ${\mathcal{M}}$:

\begin{equation}
\bigcup_{\alpha \in A} U_{\alpha} = \mathcal{M}.
\end{equation}
However, there also exists the related notion of a covering map in algebraic topology, which is a continuous surjective map between topological spaces $X$ and $C$\cite{spanier}\cite{chernavskii}:

\begin{equation}
p : C \to X,
\end{equation}
such that, for every point ${x \in X}$, there exists an open neighborhood of $x$, denoted ${U \subset X}$, such that the pre-image of $U$ under the map $p$, i.e. ${p^{-1} \left( U \right)}$, is a union of disjoint open sets in $C$, and such that $p$ maps each such open set homeomorphically onto $U$. Here, $C$ is known as the total space (or covering space) of the covering, and $X$ is known as the base space.

If the covering space $C$ is simply connected, then it is known as a universal covering space, since it covers any connected cover of the base space $X$\cite{munkres}. More precisely, if ${p : C \to X}$ is a universal cover of $X$, and if ${q : D \to X}$ is an arbitrary cover of $X$, then there must exist a covering map ${f : C \to D}$ such that ${q \circ f = p}$. In particular, in the case where the base space $X$ is a Riemannian manifold, then so too is its universal cover $C$, and the map $p$ becomes a local isometry. Therefore, one is able to generalize the concept of a Riemannian manifold to allow for connected components with variable dimensionality in the following way: one fixes a base space of a fixed dimensionality, and then allows for a universal covering space of variable dimensionality, with the universal cover of the space playing the role of the coordinate chart of the manifold. Thus, the ``curvature'' of such a generalized manifold becomes a measure of the discrepancy between the local dimensionality of the universal covering space and the global dimensionality of the base space. In the discrete case of (directed) hypergraphs, the base space can be considered to be a group-theoretical lattice, produced by some ``modding-out'' procedure. Then, the relationship between the discrete generalized manifold and the underlying base space is directly analogous to the relationship between a Bethe lattice\cite{bethe}\cite{baxter} and its underlying Cayley graph\cite{ostilli} in geometric group theory.

More specifically, if $G$ is a group and $S$ is its generating set, then the Cayley graph of $G$, denoted ${\Gamma = \Gamma \left( G, S \right)}$, is the colored directed graph whose vertex set is given by the elements of $G$\cite{cayley}\cite{magnus}:

\begin{equation}
V \left( \Gamma \right) = \lbrace g : g \in G \rbrace,
\end{equation}
and whose edge set is given by:

\begin{equation}
E \left( \Gamma \right) = \lbrace \left( g, g s \right) : g \in G, s \in S \rbrace,
\end{equation}
where the element ${s \in S}$ determines the color of the edge. On the other hand, the Bethe lattice for a given coordination number, ${z \in \mathbb{N}}$, is an infinite, connected, acyclic graph (i.e. a rooted tree), in which each vertex has exactly $z$ neighbors, such that the number of vertices contained within the $k$th shell surrounding the root vertex is given by:

\begin{equation}
N_k = z \left( z - 1 \right)^{k - 1}, \qquad k > 0.
\end{equation}
In the context of geometric group theory, the Bethe lattice with coordination number ${z = 2n}$ may be considered to be the Cayley graph of the free group on $n$ generators. Since any presentation of a group $G$ by a set $S$ of $n$ generators can be thought of as a surjective map from the free group on $n$ generators to the group $G$, all group presentations are effectively maps from the Bethe lattice (with the root vertex corresponding to the identity element) to the associated Cayley graph. In algebraic topology, this map is interpreted as the universal cover of the Cayley graph (i.e. the base space), where the Bethe lattice (the universal covering space) is guaranteed to be simply connected, even though the corresponding Cayley graph is not. More complete mathematical details of these variable-dimensionality ``pseudo-manifolds'', their connections to geometric group theory, and the possibility of formulating an analog of general relativity in which dimension is treated as a dynamical variable, will be presented in a subsequent publication.

In contemporary physics, inflationary cosmology is a modification to the standard model (otherwise known as the ${\Lambda CDM}$ or ``hot big bang'' model) of cosmology, invoked in order to explain the observed large-scale homogeneity and isotropy of the universe, and hence to solve the so-called ``horizon'' and ``flatness'' problems that are endemic to ${\Lambda CDM}$\cite{peacock}\cite{dicke}\cite{guth}\cite{fixsen}. Here, we sketch briefly how the Wolfram Model, when allowing for both local and global variation in dimensionality, is compatible with a model of cosmology which yields identical observational consequences to inflation, and hence also constitutes a valid solution to the horizon and flatness problems.

We begin by assuming that the initial condition for the universe consists of a spatial hypergraph with an abnormally high vertex connectivity, perhaps corresponding to a complete graph. As such, the universe starts off with some arbitrarily large number of spatial dimensions (which we can treat as being effectively infinite), but then the asymptotic dimensionality preserving property of the update rules causes the number of spatial dimensions to converge to some finite, fixed value, such as three. Another way to view such a universe, from the point of view of its causal structure, is that it starts with an arbitrary large value of the speed of light (since the causal graph is arbitrarily densely connected, yielding an abnormally high effective maximum rate of information propagation in the hypergraph), which then converges to a some much lower, fixed value. In this way, we can make an explicit mathematical connection with so-called ``Variable Speed of Light'' (or VSL) models of cosmology, as investigated by Petit, Moffat, Albrecht and Magueijo\cite{petit}\cite{petit2}\cite{petit3}\cite{moffat}\cite{moffat2}\cite{albrecht}, many of which are known to yield similar observational consequences to more conventional inflationary models.

An elementary mathematical model of VSL cosmology uses a scalar field, denoted ${\chi}$, to describe the dynamical behavior of the effective speed of light:

\begin{equation}
c \left( t \right) = \hat{c} \chi \left( t \right),
\end{equation}
where ${\hat{c}}$ is a constant with the dimensions of velocity. In the simplest case, we can consider a phase transition in the effective speed of light at absolute time ${t \sim t_c}$, given by:

\begin{equation}
c \left( t \right) = c_0 \theta \left( t_c - t \right) + c_m \theta \left( t - t_c \right),
\end{equation}
where ${c_m}$ is the current (measured) value of the speed of light, ${c_0 \gg c_m}$ is the initial value of the speed of light, and ${\theta \left( t \right)}$ is the Heaviside step function:

\begin{align}
\theta \left( t \right) = \begin{cases}
1, \qquad & \text{ if } t > 0,\\
0, \qquad & \text{ if } t \leq 0.
\end{cases}
\end{align}
As such, the standard FLRW (Friedmann-Lema\^itre-Robertson-Walker) metric for an expanding universe, defined by the spacetime line element:

\begin{equation}
d s^2 = g_{\mu \nu} d x^{\mu} d x^{\nu} = d t^2 c^2 \left( t \right) - a^2 \left( t \right) \left[ \frac{d r^2}{1 - k r^2} + r^2 \left( d \theta^2 + \sin \left( \theta \right) d \phi^2 \right) \right],
\end{equation}
with scale factor ${a \left( t \right)}$, now takes on a bimetric form:

\begin{equation}
g_{\mu \nu} = g_{{0}_{\mu \nu}} + g_{{m}_{\mu \nu}},
\end{equation}
where:

\begin{equation}
d s_{0}^{2} = g_{{0}_{\mu \nu}} d x^{\mu} d x^{\nu} = d t^2 c_{0}^{2} \theta \left( t_c - t \right) - a^2 \left( t \right) \left[ \frac{d r^2}{1 - k r^2} + r^2 \left( d \theta^2 + \sin^2 \left( \theta \right) d \phi^2 \right) \right],
\end{equation}
and:

\begin{equation}
d s_{m}^{2} = g_{{m}_{\mu \nu}} d x^{\mu} d x^{\nu} = d t^2 c_{m}^{2} \theta \left( t - t_c \right) - a^2 \left( t \right) \left[ \frac{d r^2}{1 - k r^2} + r^2 \left( d \theta^2 + \sin^2 \left( \theta \right) d \phi^2 \right) \right],
\end{equation}
thus producing two distinct light cones, ${d s_{0}^{2} = 0}$ and ${d s_{m}^{2} = 0}$, with the dimensionless ratio:

\begin{equation}
\gamma = \frac{c_0}{c_m},
\end{equation}
determining their relative sizes.

To see how this yields a valid solution to the horizon problem, one can compute the proper horizon scale using:

\begin{equation}
d_H \left( t \right) = a \left( t \right) \int_{0}^{t} \frac{c \left( t^{\prime} \right) d t^{\prime}}{a \left( t^{\prime} \right)},
\end{equation}
with ${t^{\prime}}$ defined by the diffeomorphism:

\begin{equation}
d t^{\prime} = a \left( t \right) d t.
\end{equation}
Since, in a radiation-dominated universe, ${a \left( t \right) \sim t^{\frac{1}{2}}}$, one obtains the standard result ${d_H \sim 2 c_m t}$, when ${t > t_c}$ and ${c = c_m}$, and ${d_h \sim 2 c_0 t}$ when ${t \sim t_c}$ and ${c = c_0}$. Therefore, when ${\gamma \to \infty}$, the size of the proper horizon becomes arbitrarily large, hence allowing all observers to be in causal contact. In particular, the forward light cone from the initial singularity becomes much larger than the region from which cosmic microwave background photons are currently being observed, as required for isotropy of the horizon.

On the other hand, the solution to the flatness problem can be seen by inspecting the Friedmann equation:

\begin{equation}
H^2 \left( t \right) + \frac{c^2 k}{a^2 \left( t \right)} = \frac{8 \pi G \rho}{3} + \frac{c^2 \Lambda}{3},
\end{equation}
where the Hubble parameter, denoted $H$, is given by:

\begin{equation}
H \left( t \right) = \frac{\dot{a} \left( t \right)}{a \left( t \right)}.
\end{equation}
Assuming zero cosmological constant, ${\Lambda = 0}$, and defining:

\begin{equation}
\epsilon \left( t \right) = \lvert \Omega \left( t \right) - 1 \rvert = \frac{c^2 \lvert k \rvert}{\dot{a}^2 \left( t \right)},
\end{equation}
where:

\begin{equation}
\Omega \left( t \right) = \frac{8 \pi G \rho}{3 H^2 \left( t \right)},
\end{equation}
we obtain:

\begin{equation}
\dot{\epsilon} \left( t \right) = - \frac{2 c^2 \lvert k \rvert \ddot{a} \left( t \right)}{\dot{a}^3 \left( t \right)} + 2 \left( \frac{\dot{c}}{c} \right) \left( \frac{c^2 \lvert k \rvert}{\dot{a}^2 \left( t \right)} \right).
\end{equation}
In a radiation-dominated universe, ${\ddot{a} \left( t \right) < 0}$, and when the speed of light decreases from ${c_0}$ to ${c_m}$, one has ${\frac{\dot{c}}{c} < 0}$, and therefore ${\dot{\epsilon} \left( t \right) < 0}$. This corresponds to a cosmological attractor solution, with ${\epsilon \sim 0}$, i.e. an approximately spatially-flat universe, as required.

This connection between higher-dimensional (i.e. more densely connected) initial spatial hypergraphs and the exponential expansion rates associated with inflationary cosmology should not be entirely surprising, since even within standard FLRW cosmology it is known that higher values for the dimensionality of space yield higher initial expansion rates\cite{lima}. This can be seen explicitly by considering the FLRW spacetime line element in $n$ spatial dimensions\cite{tangherlini}:

\begin{equation}
d s^2 = d t^2 - a^2 \left( t \right) \left( 1 + \frac{k r^2}{4} \right)^{-1} \delta_{i j} d x^i d x^j,
\end{equation}
for ${i, j = 1, 2, 3, \dots, n}$, and:

\begin{equation}
r^2 = \sum_{i} \left( x_i \right)^2.
\end{equation}
For the case of a perfect relativistic fluid, with energy density ${\rho}$ and fluid pressure $p$, obeying the higher-dimensional energy conservation law:

\begin{equation}
\dot{\rho} + n \left( \rho + p \right) \frac{\dot{a}}{a} = 0,
\end{equation}
this yields a reduced form for the higher-dimensional Einstein field equations, with $n$-dimensional gravitational constant ${G_n}$, namely:

\begin{equation}
\frac{n (n - 1)}{2} \left[ \frac{\dot{a}^2}{a^2} + \frac{k}{a^2} \right] = 8 \pi G_n \rho,
\end{equation}
and:

\begin{equation}
\frac{(n - 1) \ddot{a}}{a} + \frac{(n - 1) (n - 2)}{2} \left[ \frac{\dot{a}^2}{a^2} + \frac{k}{a^2} \right] = - 8 \pi G_n p.
\end{equation}
For a spatially flat universe ${k = 0}$, with a non-zero cosmological constant ${\Lambda}$, and a higher-dimensional cosmological equation of state parameter ${\omega_n}$\cite{turner}:

\begin{equation}
p = \omega_n \rho,
\end{equation}
the expansion term in the FLRW equations therefore takes the following analytic form, as derived by Holanda and Pereira\cite{holanda}:

\begin{equation}
a(t) = \left( \frac{n (n - 1)}{8 \Lambda} \right)^{\frac{1}{n \left( \omega_n + 1 \right)}} \frac{\left[ \exp \left( \left( \omega_n + 1 \right) \sqrt{\frac{2 \Lambda n}{n - 1}} t \right) - 1 \right]^{\frac{2}{n \left( \omega_n + 1 \right)}}}{\exp \left( \sqrt{\frac{2 \Lambda}{n (n - 1)}} t \right)},
\end{equation}
thus implying higher early-time expansion rates and lower late-time expansion rates for increasing values of $n$, and consequently allowing one to reproduce the effects of initially exponential expansion rates from cosmic inflation with arbitrarily high-dimensional initial spatial hypergraphs.

\section{Concluding Remarks}

The present article has demonstrated the Wolfram Model to be a novel, exciting and potentially highly fruitful discrete model for spacetime geometry, exhibiting discrete analogs of many (and possibly all) of the salient mathematical features of Lorentzian and pseudo-Riemannian manifolds in limiting cases. There exist a variety of open problems arising from this work, ranging from the possibility of computing higher-order corrections to the discrete Einstein field equations, to determining the computability-theoretic and complexity-theoretic properties that distinguish inertial and non-inertial reference frames, to developing a theory of general relativity that holds in manifolds with variable spacetime dimensions. A few of these problems are discussed in greater depth in our accompanying publication on quantum mechanics\cite{gorard_new}, which makes significant use of both the special relativistic and general relativistic formalisms that we develop in this paper (especially the relationship between confluence, causal invariance and Lorentz covariance, and the derivation of the discrete Einstein field equations), and we intend to investigate several more of these questions in the course of future publications. The present work, however, has at least revealed the Wolfram Model to be a plausible fundamental model for classical relativistic and gravitational physics, and we eagerly await the implications that this will entail.

\section*{Acknowledgments}

The author is greatly indebted to the other members of the Wolfram Physics Project, namely Stephen Wolfram and Max Piskunov, for their many and various contributions and insights relating to the present work. SW is responsible for originally proposing the formalism of the Wolfram Model, for determining many of its salient empirical and conceptual properties, for contributing greatly to the development of the intellectual framework of which this work is a part, and (most notably through the writing of \textit{A New Kind of Science}) for inventing and continuing to champion the very scientific paradigm within which this work was conducted. MP is responsible for much of the software development work that has spearheaded the present project, and in particular remains the principal developer of the \textit{SetReplace} paclet\cite{piskunov}, used (among other things) to render the images in Section \ref{sec:CurrentStatus0} of the present text.

The author would also like to thank Isabella Retter and Ed Pegg Jr. for useful conversations (regarding discrete differential geometry and combinatorial enumeration, respectively), {\O}yvind Tafjord and Todd Rowland for their ample supply of NKS-related source material, and Sushma Kini for keeping us all on speaking terms.


\begin{thebibliography}{99}

\bibitem{wolfram}
S. Wolfram (2002), ``A New Kind of Science''. Wolfram Media, Inc. ISBN 1-57955-008-8. \url{https://www.wolframscience.com/nks/}

\bibitem{wolfram2}
S. Wolfram (1983), ``Statistical Mechanics of Cellular Automata''. \textit{Reviews of Modern Physics}, \textbf{55} (3): 601-644. \url{https://www.stephenwolfram.com/publications/cellular-automata-complexity/pdfs/statistical-mechanics-cellular-automata.pdf}

\bibitem{wolfram3}
S. Wolfram (1985), ``Cryptography with Cellular Automata''. \textit{Proceedings of Advances in Cryptology - CRYPTO '85}: 429. \url{https://www.stephenwolfram.com/publications/academic/cryptography-cellular-automata.pdf}

\bibitem{cook}
M. Cook (2004), ``Universality in Elementary Cellular Automata'', \textit{Complex Systems}, \textbf{15}: 1-40. \url{https://wpmedia.wolfram.com/uploads/sites/13/2018/02/15-1-1.pdf}

\bibitem{cook2}
M. Cook (2008), ``A Concrete View of Rule 110 Computation'', \textit{The Complexity of Simple Programs. Electronic Proceedings in Theoretical Computer Science}, \textbf{1}: 31-55. \url{https://arxiv.org/pdf/0906.3248.pdf}

\bibitem{wolfram4}
S. Wolfram (2015), ``What is Spacetime Really?''. \url{https://writings.stephenwolfram.com/2015/12/what-is-spacetime-really/}

\bibitem{wolfram5}
S. Wolfram (1985), ``Undecidability and Intractability in Theoretical Physics'', \textit{Physical Review Letters}, \textbf{54} (8): 735-738. \url{https://www.stephenwolfram.com/publications/academic/undecidability-intractability-theoretical-physics.pdf}

\bibitem{gorard}
J. Gorard (2018), ``The Slowdown Theorem: A Lower Bound for Computational Irreducibility in Physical Systems'', \textit{Complex Systems}, \textbf{27} (2): 177-185. \url{https://wpmedia.wolfram.com/uploads/sites/13/2018/09/27-2-5.pdf}

\bibitem{shah}
R. Shah, J. Gorard (2019), ``Quantum Cellular Automata, Black Hole Thermodynamics and the Laws of Quantum Complexity'', \textit{Complex Systems}, \textbf{28} (4): 393-409. \url{https://content.wolfram.com/uploads/sites/13/2019/12/28-4-1.pdf}

\bibitem{wolfram_new}
S. Wolfram (2020), ``A Class of Models With Potential to Represent Fundamental Physics''. \url{https://www.wolframphysics.org/technical-introduction/}

\bibitem{berge}
C. Berge (1989), ``Hypergraphs: Combinatorics of Finite Sets''. North-Holland Publishing Company. ISBN 978-0444548887.

\bibitem{loll}
R. Loll (2001), ``Discrete Lorentzian Quantum Gravity'', \textit{Nuclear Physics B - Proceedings Supplements}, \textbf{94}: 96-107. \url{https://arxiv.org/pdf/hep-th/0011194.pdf}

\bibitem{ambjorn}
J. Ambj{\o}rn, A. Dasgupta, J. Jurkiewicz, R. Loll (2002), ``A Lorentzian Cure for Euclidean Troubles'', \textit{Nuclear Physics B - Proceedings Supplements}, \textbf{106}: 977-979. \url{https://arxiv.org/pdf/hep-th/0201104.pdf}

\bibitem{loll2}
R. Loll, J. Ambj{\o}rn, J. Jurkiewicz (2006), ``The Universe from Scratch'', \textit{Contemporary Physics}, \textbf{47}: 103-116. \url{https://arxiv.org/pdf/hep-th/0509010.pdf}

\bibitem{ambjorn2}
J. Ambj{\o}rn, J. Jurkiewicz, R. Loll (2005), ``Reconstructing the Universe'', \textit{Physical Review D}, \textbf{72} (064014). \url{https://arxiv.org/pdf/hep-th/0505154.pdf}

\bibitem{baader}
F. Baader, T. Nipkow (1998), ``Term Rewriting and All That''. Cambridge University Press. ISBN 978-0521779203.

\bibitem{bezem}
M. Bezem, J. Willem Klop, R. de Vrijer (2003), ``Term Rewriting Systems (`Terese')''. Cambridge University Press. ISBN 0-521-39115-6.

\bibitem{stokkermans}
K. Stokkermans (1995), ``Towards a Categorical Calculus for Critical-Pair/Completion'', \textit{Automated Practical Reasoning: Texts and Monographs in Symbolic Computation (A Series of the Research Institute for Symbolic Computation)}, 91-124. Springer, Vienna, ISBN 978-3-211-82600-3.

\bibitem{gorard_new}
J. Gorard (2020), ``Some Quantum Mechanical Properties of the Wolfram Model''. \url{https://www.wolframcloud.com/obj/wolframphysics/Documents/some-quantum-mechanical-properties-of-the-wolfram-model.pdf}

\bibitem{church}
A. Church, J. Barkley Rosser (1936), ``Some Properties of Conversion'', \textit{Transactions of the American Mathematical Society}, \textbf{39} (3): 472-482. \url{https://www.cs.cmu.edu/~crary/819-f09/ChurchRosser36.pdf}

\bibitem{battista}
G. di Battista, P. Eades, R. Tamassia, I. G. Tollis (1998), ``Flow and Upward Planarity'', \textit{Graph Drawing: Algorithms for the Visualization of Graphs}: 171-213. Prentice Hall. ISBN 978-0-13-301615-4.

\bibitem{battista2}
G. di Battista, F. Frati (2012), ``Drawing Trees, Outerplanar Graphs, Series-Parallel Graphs, and Planar Graphs in Small Area'', \textit{Thirty Essays on Geometric Graph Theory}, Algorithms and Combinatorics, Springer, \textbf{29}: 121-165.

\bibitem{naber}
G. L. Naber (1992), ``The Geometry of Minkowski Spacetime'', New York: Springer-Verlag. ISBN 978-0-387-97848-2.

\bibitem{arnowitt}
R. Arnowitt, S. Deser, C. Misner (1959), ``Dynamical Structure and Definition of Energy in General Relativity'', \textit{Physical Review}, \textbf{116} (5): 1322-1330. \url{https://authors.library.caltech.edu/72877/1/PhysRev.116.1322.pdf}

\bibitem{misner}
C. W. Misner, K. S. Thorne, J. A. Wheeler (1973), ``Gravitation''. W. H. Freeman. ISBN 978-0-7167-0344-0.

\bibitem{penrose}
R. Penrose (1972), ``Techniques of Differential Topology in Relativity''. Society for Industrial and Applied Mathematics (SIAM). ISBN 0898710057.

\bibitem{levichev}
A. V. Levichev (1987), ``Prescribing the Conformal Geometry of a Lorentz Manifold by Means of its Causal Structure'', \textit{Soviet Mathematics Doklady}, \textbf{35}: 452-455. \url{https://link.springer.com/content/pdf/10.1007%2FBF00970346.pdf}

\bibitem{malament}
D. Malament (1977), ``The Class of Continuous Timelike Curves Determines the Topology of Spacetime'', \textit{Journal of Mathematical Physics}, \textbf{18} (7): 1399-1404. \url{https://aip.scitation.org/doi/pdf/10.1063/1.523436?class=pdf}

\bibitem{welch}
P. D. Welch (2008), ``The Extent of Computation in Malament-Hogarth Spacetimes'', \textit{The British Journal for the Philosophy of Science}, \textbf{59} (4): 659-674. \url{https://people.maths.bris.ac.uk/~mapdw/borel2.1.pdf}

\bibitem{hogarth}
M. Hogarth (1992), ``Does general relativity allow an observer to view an eternity in a finite time?'', \textit{Foundations of Physics Letters}, \textbf{5} (2): 173-181.

\bibitem{hogarth2}
M. Hogarth (1994), ``Non-Turing computers and non-Turing computability'', \textit{PSA: Proceedings of the Biennial Meetings of the Philosophy of Science Association, Volume 1}: 126-138.

\bibitem{hogarth3}
M. Hogarth (2004), ``Deciding arithmetic using SAD computers'', \textit{British Journal for the Philosophy of Science}, \textbf{55} (1): 681-991.

\bibitem{kuratowski}
K. Kuratowski (1930), ``Sur le probl\`eme des courbes gauches en topologie'', \textit{Fundamenta Mathematicae}, \textbf{15}: 271-283. \url{http://matwbn.icm.edu.pl/ksiazki/fm/fm15/fm15126.pdf}

\bibitem{tutte}
W. T. Tutte (1963), ``How to Draw a Graph'', \textit{Proceedings of the London Mathematical Society}, Third Series, \textbf{13}: 743-767. \url{http://www.cs.jhu.edu/~misha/Fall07/Papers/Tutte63.pdf}

\bibitem{wagner}
K. Wagner (1937), ``\"Uber eine Eigenschaft der Ebenen Komplexe'', \textit{Mathematische Annalen}, \textbf{114}: 570-590. \url{https://link.springer.com/content/pdf/10.1007%2FBF01594196.pdf}

\bibitem{robertson}
N. Robertson, P. Seymour (1983), ``Graph Minors. I. Excluding a Forest'', \textit{Journal of Combinatorial Theory, Series B}, \textbf{35} (1): 39-61. \url{https://core.ac.uk/download/pdf/82390181.pdf}

\bibitem{robertson2}
N. Robertson, P. Seymour (2004), ``Graph Minors. XX. Wagner's Conjecture'', \textit{Journal of Combinatorial Theory, Series B}, \textbf{92} (2): 325-357. \url{https://web.math.princeton.edu/~pds/papers/GM20/GM20.ps}

\bibitem{bruijn}
N. G. de Bruijn (1972), ``Lambda Calculus Notation with Nameless Dummies: A Tool for Modern Formula Manipulation, with Application to the Church-rosser Theorem'', \textit{Indagationes Mathematicae}, \textbf{34}: 381-392. ISSN 0019-3577. \url{https://www.win.tue.nl/automath/archive/pdf/aut029.pdf}

\bibitem{gabbay}
M. J. Gabbay, A. M. Pitts (1999), ``A New Approach to Abstract Syntax Involving Binders'', \textit{14th Annual IEEE Symposium on Logic in Computer Science}: 214-224. \url{https://www.cl.cam.ac.uk/~amp12/papers/newaas/newaas.pdf}

\bibitem{ricci}
G. Ricci (1904), ``Direzioni e Invarianti Principali in una Variet\`a Qualunque'', \textit{Atti del Reale Istituto veneto di Scienze}, \textbf{63} (2): 1233-1239.

\bibitem{ollivier}
Y. Ollivier (2007), ``Ricci Curvature of Metric Spaces'', \textit{Comptes Rendus Math\'ematique de l'Acad\'emie des Sciences}, \textbf{345}: 643-646. \url{https://www.math.uchicago.edu/~shmuel/QuantCourse%20/Metric%20Space/Ollivier,%20Ricci%20curvature%20of%20Metric%20Spaces.pdf}

\bibitem{ollivier2}
Y. Ollivier (2009), ``Ricci Curvature of Markov Chains on Metric Spaces'', \textit{Journal of Functional Analysis}, \textbf{256} (3): 810-864. \url{https://arxiv.org/pdf/math/0701886.pdf}

\bibitem{ollivier3}
Y. Ollivier (2011), ``A Visual Introduction to Riemannian Curvatures and Some Discrete Generalizations'', \textit{Analysis and Geometry of Metric Measure Spaces: Lecture Notes of the 50th S\'eminaire de Math\'ematiques Sup\'erieures (SMS), Montr\'eal}, \textbf{56}: 197-219. \url{http://www.yann-ollivier.org/rech/publs/visualcurvature.pdf}

\bibitem{eidi}
M. Eidi, J. Jost (2019), ``Ollivier Ricci Curvature of Directed Hypergraphs''. \url{https://arxiv.org/pdf/1907.04727.pdf}

\bibitem{hilbert}
D. Hilbert (1924), ``Die Grundlagen der Physik'', \textit{Mathematische Annalen}, \textbf{92}: 1-32. \url{https://link.springer.com/content/pdf/10.1007/BF01448427.pdf}

\bibitem{chapman}
S. Chapman, T. G. Cowling (1970). ``The Mathematical Theory of Non-Uniform Gases'' (3rd Edition). Cambridge University Press. ISBN 978-0521408448.

\bibitem{balescu}
R. Balescu (1975), ``Equilibrium and Nonequilibrium Statistical Mechanics''. John Wiley \& Sons. ISBN 978-0-471-04600-4.

\bibitem{wolfram7}
S. Wolfram (1986), ``Cellular Automaton Fluids 1: Basic Theory'', \textit{Journal of Statistical Physics}, \textbf{45} (3/4): 471-526. \url{https://www.stephenwolfram.com/publications/academic/cellular-automaton-fluids-theory.pdf}

\bibitem{cercignani}
C. Cercignani (1975), ``Theory and Application of the Boltzmann Equation''. Elsevier. ISBN 978-0-444-19450-3.

\bibitem{wolfram6}
S. Wolfram (2019), \textit{Personal Communication}.

\bibitem{milgrom}
M. Milgrom (1983), ``A Modification of the Newtonian Dynamics as a Possible Alternative to the Hidden Mass Hypothesis'', \textit{Astrophysical Journal}, \textbf{270}: 365-370. \url{http://articles.adsabs.harvard.edu/pdf/1983ApJ...270..365M}

\bibitem{lovelock}
D. Lovelock (1971), ``The Einstein Tensor and its Generalizations'', \textit{Journal of Mathematical Physics}, \textbf{12}: 498-501. \url{https://aip.scitation.org/doi/pdf/10.1063/1.1665613?class=pdf}

\bibitem{woolliams}
S. A. Woolliams (2013), ``Higher Derivative Theories of Gravity''. \url{https://www.imperial.ac.uk/media/imperial-college/research-centres-and-groups/theoretical-physics/msc/dissertations/2013/LI_INCOMPACT3D2014.pdf}

\bibitem{hawking}
S. W. Hawking, G. F. R. Ellis (1973), ``The Large Scale Structure of Space-Time''. Cambridge University Press. ISBN 0-521-09906-4.

\bibitem{danehkar}
A. Danehkar (2009), ``On the Significance of the Weyl Curvature in a Relativistic Cosmological Model'', 
\textit{Modern Physics Letters A}, \textbf{24} (38): 3113-3127. \url{https://arxiv.org/pdf/0707.2987.pdf}

\bibitem{lee}
J. Lee (2010), ``Introduction to Topological Manifolds''. Springer Science \& Business Media. ISBN 978-1-4419-7940-7.

\bibitem{nottale}
L. Nottale (1989), ``Fractals and the Quantum Theory of Spacetime'', \textit{International Journal of Modern Physics A}, \textbf{04} (19): 5047-5117. \url{https://www.worldscientific.com/doi/pdf/10.1142/S0217751X89002156}

\bibitem{ord}
G. N. Ord (1983), ``Fractal Space-Time: A Geometric Analogue of Relativistic Quantum Mechanics'', \textit{Journal of Physics A: Mathematical and General}, \textbf{16} (9): 1869-1884. \url{https://iopscience.iop.org/article/10.1088/0305-4470/16/9/012/pdf}

\bibitem{nottale2}
L. Nottale, J. Schneider (1984), ``Fractals and Nonstandard Analysis'', \textit{Journal of Mathematical Physics}, \textbf{25} (5): 1296-1300. \url{https://aip.scitation.org/doi/pdf/10.1063/1.526285}

\bibitem{jost}
J. Jost (2013), ``Riemannian Geometry and Geometric Analysis''. Springer Science \& Business Media. ISBN 978-3-319-61859-3.

\bibitem{lee2}
J. M. Lee (2006), ``Introduction to Smooth Manifolds''. Springer-Verlag New York. ISBN 978-0-387-95448-6.

\bibitem{spanier}
E. H. Spanier (1994), ``Algebraic Topology''. Springer-Verlag New York. ISBN 978-0-387-94426-5.

\bibitem{chernavskii}
A. V. Chernavskii (2001), ``Covering'', \textit{Encyclopedia of Mathematics}. Springer Science \& Business Media. ISBN 978-1-55608-010-4.

\bibitem{munkres}
J. R. Munkres (2000), ``Topology'' (2nd Edition). Upper Saddle River, New Jersey: Prentice Hall. ISBN 0131816292.

\bibitem{bethe}
H. A. Bethe (1935), ``Statistical Theory of Superlattices'', \textit{Proceedings of the Royal Society of London A}, \textbf{150}: 552-575. \url{https://royalsocietypublishing.org/doi/pdf/10.1098/rspa.1935.0122}

\bibitem{baxter}
R. J. Baxter (1982), ``Exactly Solved Models in Statistical Mechanics''. Academic Press. ISBN 0-12-083182-1.

\bibitem{ostilli}
M. Ostilli (2012), ``Cayley Trees and Bethe Lattices, a Concise Analysis for Mathematicians and Physicists'', \textit{Physica A}, \textbf{391}: 3417-3424. \url{https://arxiv.org/pdf/1109.6725.pdf}

\bibitem{cayley}
A. Cayley (1878), ``Desiderata and Suggestions: No. 2. The Theory of Groups: Graphical Representation'', \textit{American Journal of Mathematics}, \textbf{1} (2): 174-176. \url{https://www.jstor.org/stable/2369306?seq=1#metadata_info_tab_contents}

\bibitem{magnus}
W. Magnus, A. Karrass, D. Solitar (2004), ``Combinatorial Group Theory: Presentations of Groups In Terms of Generators and Relations''. Courier. ISBN 978-0-486-43830-6.

\bibitem{peacock}
J. A. Peacock (1998), ``Cosmological Physics''. Cambridge University Press. ISBN 978-0-521-42270-3.

\bibitem{dicke}
R. H. Dicke (1970), ``Gravitation and the Universe: Jayne Lectures for 1969''. American Philosophical Society. ISBN 978-0871690784.

\bibitem{guth}
A. Guth (1981), ``Inflationary Universe: A Possible Solution to the Horizon and Flatness Problems'', \textit{Physical Review D}, \textbf{23} (2): 347-356. \url{http://tucana.astro.physik.uni-potsdam.de/~cfech/lectures/galkos/guth1981.pdf}

\bibitem{fixsen}
D. J. Fixsen (2009), ``The Temperature of the Cosmic Microwave Background'', \textit{The Astrophysical Journal}, \textbf{707} (2): 916-920. \url{https://arxiv.org/pdf/0911.1955.pdf}

\bibitem{petit}
J. P. Petit (1988), ``An Interpretation of Cosmological Model with Variable Light Velocity'', \textit{Modern Physics Letters A}, \textbf{3} (16): 1527-1532. \url{https://www.jp-petit.org/science/f300/modern_physics_letters_a1.pdf}

\bibitem{petit2}
J. P. Petit (1988), ``Cosmological Model with Variable Light Velocity: The Interpretation of Red Shifts'', \textit{Modern Physics Letters A}, \textbf{3} (18): 1733-1744. \url{https://www.jp-petit.org/science/f300/modern_physics_letters_a2.pdf}

\bibitem{petit3}
J. P. Petit, M. Viton (1989), ``Gauge Cosmological Model with Variable Light Velocity. Comparison with QSO Observational Data'', \textit{Modern Physics Letters A}, \textbf{4} (23): 2201-2210. \url{https://www.jp-petit.org/science/f300/modern_physics_letters_a3.pdf}

\bibitem{moffat}
J. W. Moffat (1993), ``Superluminary Universe: A Possible Solution to the Initial Value Problem in Cosmology'', \textit{International Journal of Modern Physics D}, \textbf{2} (3): 351-366. \url{https://arxiv.org/pdf/gr-qc/9211020.pdf}

\bibitem{moffat2}
J. W. Moffat (2002), ``Variable Speed of Light Cosmology: An Alternative to Inflation''. \url{https://arxiv.org/pdf/hep-th/0208122.pdf}

\bibitem{albrecht}
A. Albrecht, J. Magueijo (1999), ``A Time Varying Speed of Light as a Solution to Cosmological Puzzles'', \textit{Physical Review D}, \textbf{59} (4): 043516. \url{https://arxiv.org/pdf/astro-ph/9811018.pdf}

\bibitem{lima}
J. A. S. Lima (2001), ``Note on solving for the dynamics of the Universe'', \textit{American Journal of Physics}, \textbf{59}: 1245-1247. \url{https://aapt.scitation.org/doi/10.1119/1.1405506}

\bibitem{tangherlini}
F. R. Tangherlini (1963), ``Schwarzschild field in $n$ dimensions and the dimensionality of space problem'', \text{Il Nuovo Cimento}, \textbf{27} (3): 636-651. \url{https://www.osti.gov/biblio/4749059-schwarzschild-field-inn-dimensions-dimensionality-space-problem}

\bibitem{turner}
M. S. Turner and M. White (1997), ``CDM models with a smooth component'', \textit{Physics Review D} \textbf{56} (8): R4439(R). \url{https://journals.aps.org/prd/abstract/10.1103/PhysRevD.56.R4439}

\bibitem{holanda}
R. F. L. Holanda and S. H. Pereira (2012), ``On the Dynamics of the Universe in $D$ Spatial Dimensions'', \textit{Revista Mexicana de Astronom\'ia y Astrof\'isica}, \textbf{48} (2): 251-256. \url{https://arxiv.org/abs/0707.3387}

\bibitem{piskunov}
M. Piskunov, \textit{SetReplace}, \url{https://github.com/maxitg/SetReplace}.
\end{thebibliography}
\end{document}